\documentclass[journal, twocolumn, 10pt]{IEEEtran}
\usepackage{url}
\usepackage{graphicx}
\usepackage{epstopdf}
\usepackage{algorithm}
\usepackage{algorithmicx}
\usepackage{algpseudocode}
\usepackage{cite}
\usepackage{color}
\usepackage{amsmath,amssymb,amsfonts,amsthm}
\usepackage{bm}
\usepackage{setspace}
\usepackage{multirow}
\usepackage{mathtools}
\usepackage{wrapfig}
\usepackage{subcaption}
\usepackage{multicol}
\usepackage{lipsum}
\usepackage{fancyhdr}

\newcommand{\lb}{\left(}
\newcommand{\rb}{\right)}

\newcommand{\bone}{\mathbf{1}}

\newcommand{\bdin}{\mathbf{d}_{\textnormal{in}}}

\newcommand{\bx}{\mathbf{x}}
\newcommand{\by}{\mathbf{y}}

\newcommand{\br}{\mathbf{r}}

\newcommand{\bz}{\mathbf{z}}

\newcommand{\be}{\mathbf{e}}

\newcommand{\bxt}{\widetilde{\mathbf{x}}}
\newcommand{\byt}{\widetilde{\mathbf{y}}}

\newcommand{\bAbet}{\mathbf{A}^{\textnormal{bet}}}
\newcommand{\bAwit}{\mathbf{A}^{\textnormal{wit}}}

\newcommand{\bSigma}{\mathbf{\Sigma}}

\newcommand{\bA}{\mathbf{A}}

\newcommand{\bV}{\mathbf{V}}

\newcommand{\bM}{\mathbf{M}}

\newcommand{\bAt}{\widetilde{\mathbf{A}}}

\newcommand{\cR}{\mathcal{R}}

\newcommand{\cE}{\mathcal{E}}

\newcommand{\TB}{\mathbb{T}}

\newcommand{\bbR}{\mathbb{R}}

\begin{document}
	
	\title{Identifying Influential Links for Event Propagation on Twitter: A Network of Networks Approach
	}
	
\author{
	Pin-Yu Chen, Chun-Chen Tu, Pai-Shun Ting, Ya-Yun Lo, Danai Koutra, and Alfred O. Hero III,~\emph{Fellow},~\emph{IEEE}
%	$^{1}$Department of Electrical Engineering and Computer Science\\
%	$^{2}$Department of Statistics\\
%	University of Michigan, 
%	Ann Arbor, MI 48109, USA \\
		\thanks{P.-Y. Chen is with the AI Foundations Group at IBM Thomas J. Watson Research Center, Yorktown Heights, NY 10598. Email: pin-yu.chen@ibm.com. P.-S. Ting, Y.-Y. Lo, D. Koutra and A. O. Hero are with the Department of Electrical Engineering and Computer Science, University of Michigan, Ann Arbor, MI 48109, USA. Email : \{paishun,yayunlo,dkoutra,hero\}@umich.edu. C.-C. Tu is with the Department of Statistics, University of Michigan, Ann Arbor, MI 48109, USA. Email : {timtu}@umich.edu.} 
		\thanks{This work was conducted while P.-Y. Chen was at the University of Michigan, and it was supported in part by the ARO grants W911NF-15-1-0479 and W911NF-12-1-0443, and the startup funding from CoE at University of Michigan.}
}

%	\author{Pin-Yu~Chen and Alfred O. Hero III,~\emph{Fellow},~\emph{IEEE}%<-this % stops a space
%		\thanks{P.-Y. Chen and A. O. Hero are with the Department of Electrical Engineering and Computer Science, University of Michigan, Ann Arbor, MI 48109, USA. Email : pinyu@umich.edu and hero@umich.edu.}
%		\thanks{This work was supported in part by the ARO grant W911NF-15-1-0479 and startup funding from CoE at University of Michigan.}
%	}
	
	\maketitle
	\thispagestyle{empty}
	
	\begin{abstract}
Patterns of event propagation in online social networks provide novel insights on the modeling and analysis of information dissemination over networks and physical systems. This paper studies the importance of follower links for event propagation on Twitter. Three recent event propagation traces are collected with the Twitter user language field being used to identify the Network of Networks (NoN) structure embedded in the Twitter follower networks. We first formulate event propagation on Twitter as an iterative state equation, and then propose an effective score function on follower links accounting for the containment of event propagation via link removals. Furthermore, we find that utilizing the NoN model can successfully identify influential follower links such that their removals lead to remarkable reduction in event propagation on Twitter follower networks. Experimental results find that the between-network follower links, though only account for a small portion of the total follower links, are crucial to event propagation on Twitter.
	\end{abstract}
	
	
\section{Introduction}
\label{sec_Intro}	
Patterns of event propagation in online social networks, physical networks, and biological networks are closely related. Examples include epidemic spread in contact networks \cite{Pastor-Satorras15epidemic_review,nowzari2016analysis},
information diffusion in social networks and social media \cite{yang2010predicting,kitsak2010identification,myers2012information,Arruda14,kimura2009blocking,del2016spreading,Radicchi16influential}, and malware propagation in technological networks \cite{Cohen03immu,zou2007modeling,Chen08immu,gao2013modeling}, among others.

This paper exploits the network structure embedded in online social networks to identify influential links for event propagation. Specifically, we use Twitter follower networks to study and develop an effective link score function that reflects the importance of a follower link in event propagation. An event on a Twitter follower network can be a uniform resource locator (URL) of a web address or a hashtag in a tweet. 
A follower who has seen a tweet and decided (not) to retweet the event is called a retweeter (non-retweeter).
A typical example of event propagation on Twitter is the announcement of the discovery of a Higgs boson-like particle in July 2012 \cite{de2013anatomy}. Given a Twitter follower network, our proposed method effectively identifies important follower links affecting event propagation based on the network connectivity structure without requiring prior knowledge such as where the event is originally posted and how the event is retweeted.

We model event propagation using an iterative state equation, and then propose a Left Eigenvector Score (LES) for evaluating the influence of event propagation for each follower link. We show that LES is able to identify influential follower links for event propagation in the sense that the removal of those links is effective in reducing event propagation. Although our method requires only the information of the network's connectivity structure, it can be easily extended to incorporate additional user information to further improve the effectiveness of the proposed method. Specifically, we utilize the Network of Networks (NoN) structure in Twitter follower networks as additional user information. The NoN model is a general approach for characterizing a network at different scales. A large-scale network is often composed of several sub-networks, and the interconnectivity and interdependency between these sub-networks are known to be crucial to information dissemination and network robustness \cite{Gao11NoN,Roni10,buldyrev2010catastrophic,Anna12,Radicchi13,ni2014inside}. The use of the NoN model enables network algorithms to exploit the interconnectivity structure at the sub-network level, and has shown to be effective in various domains, including efficient ranking algorithms in networks \cite{ni2014inside}  and cascading analysis in interconnected systems \cite{buldyrev2010catastrophic}, among others.

To validate the effectiveness of LES and the NoN structure, we created two synthetic event propagation datasets of particular NoN structure, and collected three recent event propagation traces on Twitter using the Application Programming Interface (API)\footnote{Twitter REST APIs. Available at https://dev.twitter.com/rest/public} provided by Twitter for public data retrieval. The API offers an effective platform for tracking and collecting real-world event propagation traces on Twitter at large scales. The users' languages filed on Twitter are used to build the NoN model to identify the language-related sub-networks within the Twitter follower  network under consideration. With the NoN model, we aim to study the effect of intraconnectivity  and interconnectivity of user languages on event propagation. We find that the between-network links play an important role in event propagation, as they account for information dissemination from one user language to another. Experimental results demonstrate that link removals based on LES can successfully reduce event propagation in real-world traces, especially when the between-network follower links are used for LES calculation. In particular, the success of LES-based scores in identifying influential follower links can be explained by the fact that the LES of a link is the product of eigenvector centrality of the associated user pair based on the follower network connectivity pattern, where the eigenvector centrality of a user is proportional to the sum of eigenvector centrality of his/her followers.

	The rest of this paper is organized as follows. Sec. \ref{sec_related} provides an overview of event propagation in networks. Sec. \ref{sec_NoN_structure} defines the NoN model and illustrates the NoN structure from the collected Twitter traces. Sec. \ref{sec_methodology_NoN} provides a theoretical framework for  identifying influential follower links for event propagation, including modeling event propagation via an iterative state equation, specifying a surrogate function for event propagation, and proposing a novel link score function (LES) for evaluating the importance of follower links in event propagation. Sec. \ref{sec_experiment}
	uses the synthetic event propagation datasets  and the collected Twitter traces to compare the performance of different score functions for identifying influential follower links. Finally, Sec. \ref{sec_con} concludes this paper.

The main notations used in this paper are given in Table \ref{table_notation}.

\begin{table}[h]
	\centering
	\caption{List of main notations.}
	\label{table_notation}
\begin{tabular}{|c|c|}
	\hline
	Notation             & Description                                                                                                    \\ \hline
	$n$                  & number of users                                                                                                \\ \hline
	$m$                  & number of follower links                                                                                       \\ \hline
	$\bA$                & adjacency matrix of follower links                                                                             \\ \hline
	$\bAbet$ ($\bAwit$)  & \begin{tabular}[c]{@{}c@{}}adjacency matrix of between-network \\ (within-network) follower links\end{tabular} \\ \hline
	$\lambda_{\max}(\bA)$ & largest eigenvalue of $\bA$                                                                                    \\ \hline
	$\br_t$              & binary event propagation status vector                                                                         \\ \hline
	$\TB$                & entry-wise threshold function                                                                                  \\ \hline
	$s$                  & \begin{tabular}[c]{@{}c@{}}upper bound on the total number \\ of retweeters \end{tabular}                  \\ \hline
	$\by$                & leading left eigenvector of $\bA$                                                                              \\ \hline
	$\cE_{\cR}$          & link removal set                                                                                               \\ \hline
	$q$                  & number of removed links                                                                                        \\ \hline
	$R_f$               & reachability of a score function $f$
	\\ \hline	
\end{tabular}
\end{table}

\section{Related Work}
\label{sec_related}
	Event propagation in networks has been actively studied in many different fields. In \cite{Pastor-Satorras15epidemic_review,nowzari2016analysis,Zou05,CPY14control}, event propagation is studied in the context of epidemic processes in physical and engineering networks. Each node in the network is categorized into a few states (e.g., the susceptible-infected-recovered (SIR) model) for analyzing and predicting collective behaviors, such as the emergence of epidemic spreads, or the monitoring of malware propagation. In online social networks, event propagation is studied in the context of information diffusion \cite{cha2009measurement,yang2010predicting,myers2012information,de2013anatomy}, 	influence maximization \cite{kempe2003maximizing},  influential user identification \cite{kitsak2010identification,Arruda14,kimura2009blocking}, and locating rumor sources \cite{shah2011rumors}. In signal processing, event propagation is studied in the context of diffusion estimation among agents in a network \cite{lopes2008diffusion,Gholami16}, and extracting patterns based on the diffusion of graph signals \cite{benzi2016principal,hamon2016extraction,segarra2015diffusion}.

	Many existing event propagation models, such as the SIR model \cite{Pastor-Satorras15epidemic_review} for epidemic processes, and the independent cascade model (ICM) \cite{goldenberg2001using} for 	influence maximization, assume independent  probabilistic link activation models for event propagation. For example, 
		for the SIR model, an infected node can infect a susceptible node within the contact range with some probability, or can transition to the recovered state with some probability, where these infection and recover probabilities are governed by certain parameters. Similarly,  for the ICM, each node can be probabilistically activated for further information propagation provided that one of the neighboring nodes has been activated.

	Different from these parametric event propagation models,  we formulate event propagation as an iterative state equation that is only associated with the network structure. 
	Our event propagation model then leads to a surrogate function of topological dependence of  event propagation, as is explained in Sec. \ref{subsec_surrogate}. It allows us to evaluate the importance of every link in the network without assuming any probabilistic link activation models, which is discussed in Sec. \ref{subsec_LES}. Such an event propagation model is particularly appealing for studying event propagation in online social networks such as Twitter, since the network structure (i.e., the follower connectivity pattern) can be easily obtained from the Twitter API. 
	
As will be discussed in Sec. \ref{sec_methodology_NoN}, the surrogate function for event propagation is the largest eigenvalue of the adjacency matrix, and we propose to use the leading left eigenvector of the adjacency matrix to identify and remove influential links, leading to maximal reduction in the resulting surrogate function. Similar link removal approaches have been proposed by the NetMelt algorithm \cite{tong2012gelling}, where the authors use both the leading left and right eigenvectors to identify important links for reducing the same surrogate function. Specifically, the proposed LES of a link is the product of eigenvector centrality of the associated user pair based on the follower network connectivity pattern, where the eigenvector centrality of a user is proportional to the sum of eigenvector centrality of his/her followers. On the other hand, the link score provided by the NetMelt algorithm does not have such interpretation.
A performance comparison between the proposed method and the NetMelt algorithm is given in Sec. \ref{sec_experiment}.

	\begin{figure*}[t]
		\centering
		\begin{subfigure}[b]{0.55\linewidth}
			\includegraphics[width=\textwidth]{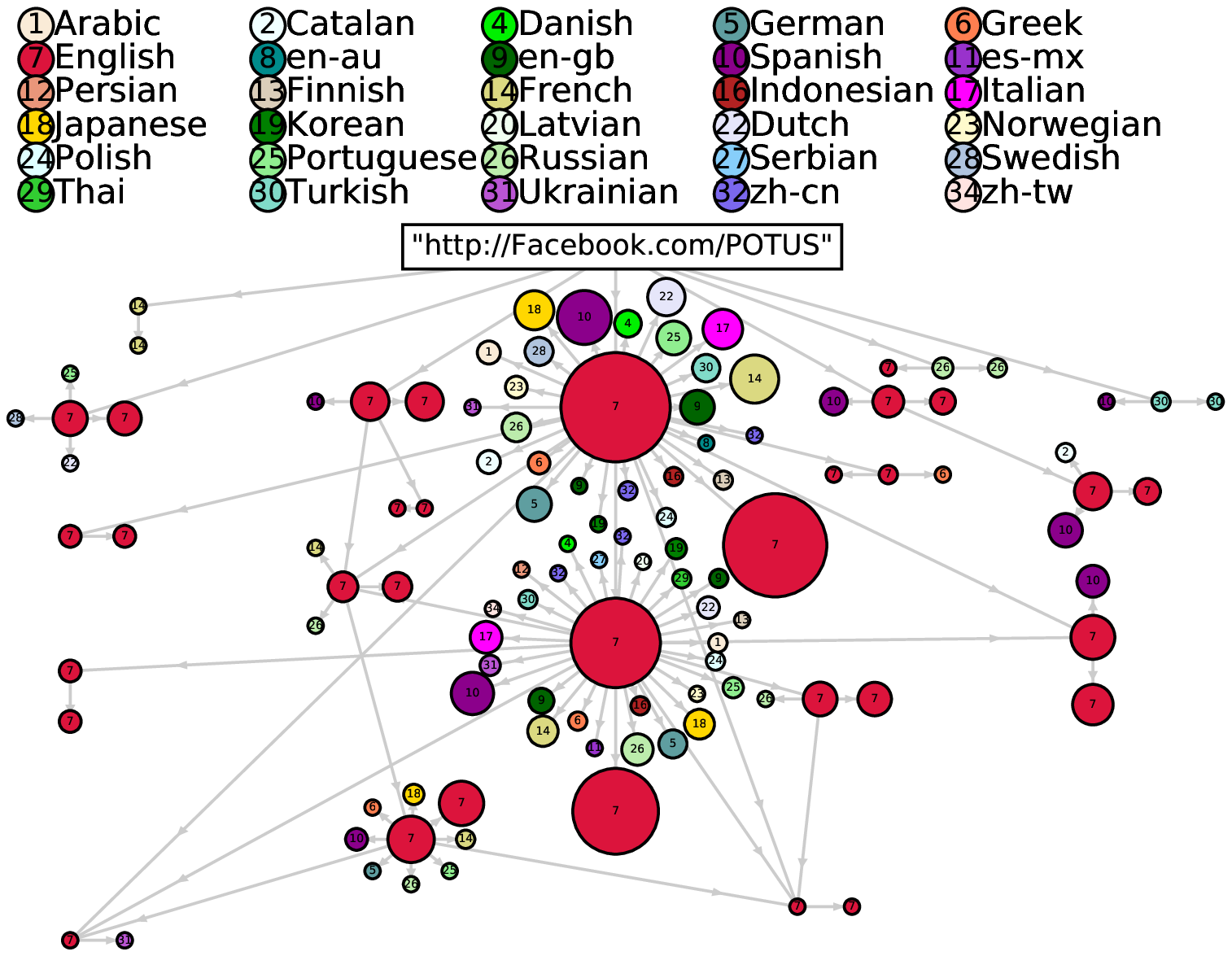}
		\vspace{-10mm}				
			\caption{Obama FB}
			%			\label{}
		\end{subfigure}%
%		\hspace{-0.1cm}
		\centering
		\begin{subfigure}[b]{0.55\linewidth}
			\includegraphics[width=\textwidth]{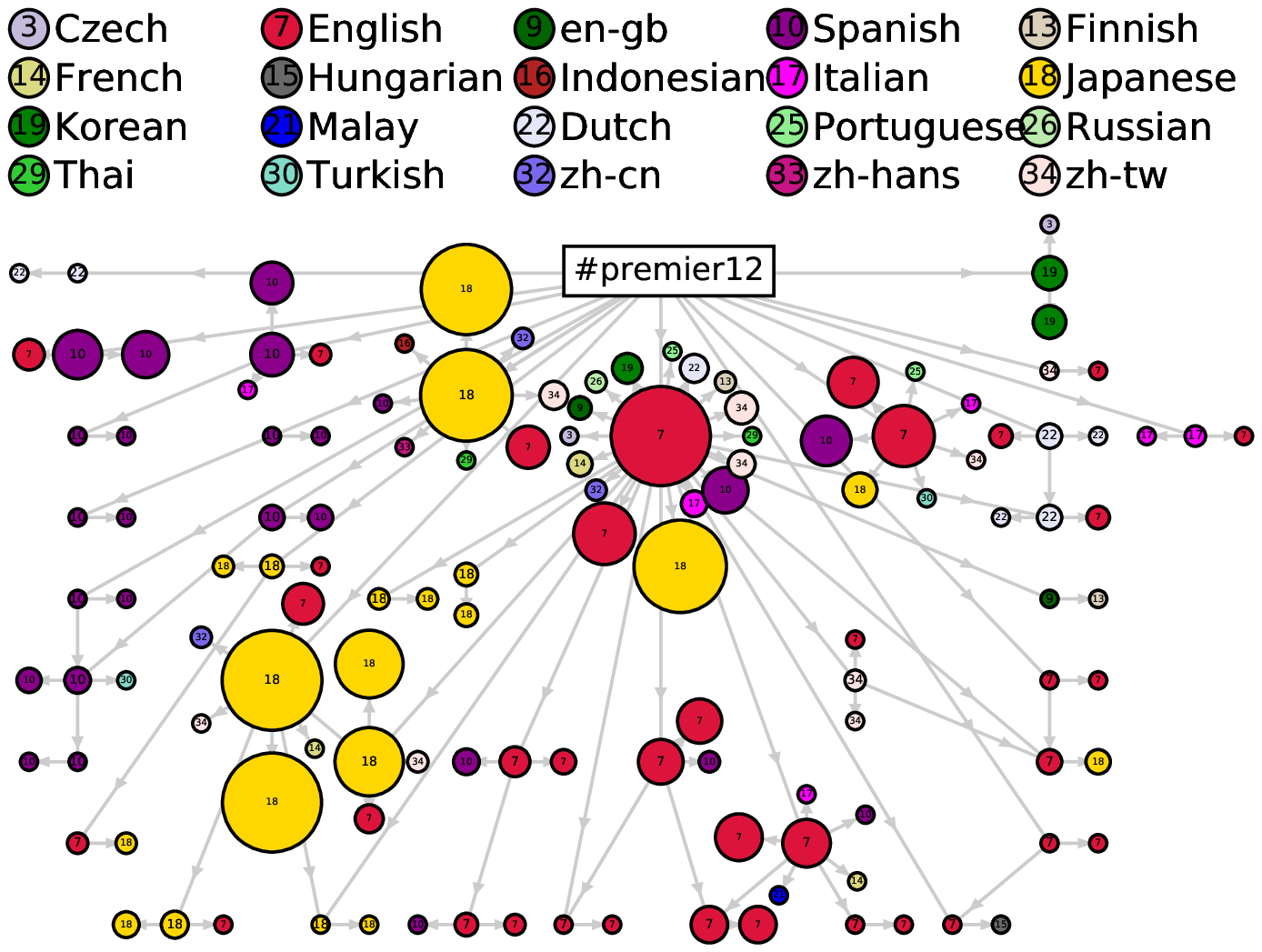}
		\vspace{-10mm}				
			\caption{Premier 12}
			%			\label{}
		\end{subfigure}
		\\
		%\vspace{3mm}
%			\hspace{-0.5cm}
		\centering
		\begin{subfigure}[b]{0.55\linewidth}
			\includegraphics[width=\textwidth]{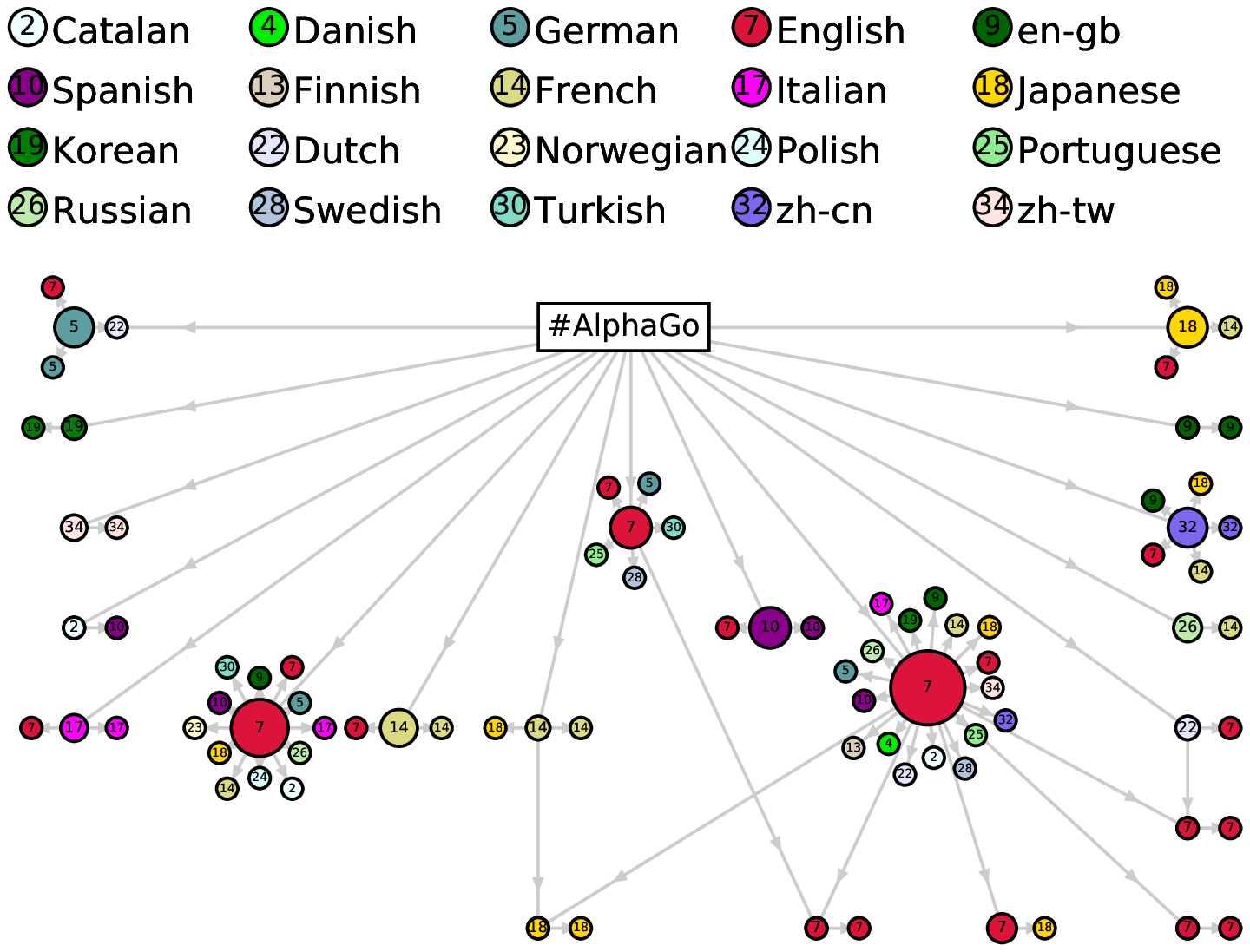}
		\vspace{-10mm}			
			\caption{AlphaGo}
			%			\label{}
		\end{subfigure}
%		\vspace*{-2mm}	
		\caption{
		The three collected retweeter networks with user language characterizing the Network of Networks (NoN) structure. 
				A retweeter is represented by a node with language setting denoted by its color/number. The direction of an edge represents the direction of event propagation.  For succinct graphical representation, we grouped all the same-language leaf retweeters of a single node into a super node. The size of a super node is proportional to the square root of the retweeter population.
				It is observed that an event is first disseminated by some seed nodes of different languages, and other nodes tend to retweet the event from a same-language node.}	
		\label{Fig_NoN_structure}
%		\vspace*{-4mm}		
	\end{figure*}

	\section{Illustration of the NoN Structure of Event Propagation on Twitter}
	\label{sec_NoN_structure}
	To concretely illustrate event propagation, we collected the traces of three recent events on Twitter during a period of two weeks through the Twitter API.  The details of the collected event traces from Twitter, including the description and collection duration, are given in Appendix \ref{appen_detail_NON}. These datasets are made public available at \url{https://sites.google.com/site/pinyuchenpage/datasets}. 
	These events include URLs and hashtags such as the following.

	\begin{itemize}
		\item \textbf{Obama FB:} A URL that links to U.S. President Obama's personal Facebook page created in 2015.
		\item \textbf{Premier 12:} A hashtag of an international baseball tournament in 2015.
		\item \textbf{AlphaGo:}	A hashtag about a board game algorithm defeating a European Go champion in 2016 \cite{silver2016mastering}. 
	\end{itemize}

The source of an event need not to be unique. For example, the same URL can be independently posted by some users and subsequently be retweeted by their followers. 

A network of networks (NoN) model of $n$ users and $L$ sub-networks is a network connectivity representation at the sub-network level, where each node (i.e., a super node) in the NoN representation refers to a sub-network that specifies a set of users and the corresponding links among them. The links within a sub-network and between sub-networks are called within-network and between-network links, respectively. To build a NoN model for the studied datasets, we collected each user's language setting on Twitter, which is used as the network identity. Specifically, for each user we use the language field as the sub-network label to represent the follower network as a network of $L$ interconnected sub-networks, where $L$ is the number of language fields. 
	The NoN model is then used  to study the role of within-network and between-network follower links on event propagation.

Figure \ref{Fig_NoN_structure} displays the NoN structure in the retweeter network of the aforementioned events. It is observed that the propagation patterns of these events share some common features.
	(i) For each event, there are some hub users such that their posts are retweeted by many other users. For the Obama FB event, one hub user is President Obama's personal Twitter account, and another hub user is White House's  Twitter account. For the Premier 12 event, one hub user is the tournament organizer's official Twitter account. For the AlphaGo event, one hub user is  Google's Twitter account. (ii) The events are originally posted by some ``seed users'' of different languages, and other users tend to retweet the event from a user of the same language. Take Premier 12 as an example, the tweets regarding Premier 12 are first tweeted by some seed users of different languages, including Dutch, English, Spanish, Korean, zh-TW and Italian. Then most of the tweets are retweeted by users of the same language.
In particular, the fraction of most-populated same-language retweeters in the Obama FB, Premier 12 and AlphaGo datasets is 80.67\%, 78.78\% and 61.27\%, respectively.
These findings suggest that the NoN model for event propagation on Twitter captures different roles of within-network and between-network follower links in event propagation. This motivates us to identify influential follower links using the NoN model. We will report significant gain in the NoN model. 	
	%At the top is the root node, labeled as node 0,  which represents the virtual source of event propagation.
	%The other numbers and colors of a node (network) represent a user sub-network of the same language. 
	%There is an edge connecting two nodes if the event has been retweeted from one network of users to another.
	%For the purpose of succinct representation, users of the same language who have retweeted an event but have not been retweeted by other users are merged into a single network.
	
	%The datasets can be downloaded from XXX.

	%Based on the NoN structure of the three events in Fig. \ref{Fig_NoN_structure}, it is observed that the propagation patterns of these events share some common features. (i) For each event, there are some hub networks such that their posts are retweeted by many other networks. For the Obama FB event, one hub network includes President Obama's personal Twitter account, and another hub network includes White House's  Twitter account. For the Premier 12 event, one hub network includes the tournament organizer's official Twitter account. For the AlphaGo event, one hub network includes Google's Twitter account. (ii) The events are originally posted by some ``seed networks'' of different languages, and other networks tend to retweet the event from a network of the same language. Take Premier 12 as an example, the tweets regarding Premier 12 are first tweeted by different seed networks, including Dutch, English, Spanish, Korean, zh-TW and Italian. Then most of the tweets are retweeted by networks of the same language.

	\section{Methodology}
	\label{sec_methodology_NoN}
	\subsection{Event propagation model}
	\label{subsec_event}
	Consider a Twitter follower network consisting of $n$ users and $m$ follower links. Let $\bA$ be an $n \times n$ binary adjacency matrix representing the follower relationship in the network, where its entry of the $i$-th row and the $j$-th column $[\bA]_{ij}=1$ if user $i$ follows user $j$, and $[\bA]_{ij}=0$ otherwise. We divide the time period of event propagation into $F$ non-overlapping frames, and let $\bA_t$ be an $n \times n$ binary adjacency matrix indicating the follower links that have been activated for event propagation during the $t$-th time frame, $t=1,2,\ldots F$. In other words, $[\bA]_{ij}=1$ indicates that user $i$ follows user $j$, while  $[\bA_t]_{ij}=1$ indicates that user $i$ retweets user $j$ during the $t$-th time frame.
	Let $\br_t$ be an $n$-dimensional binary vector indicating the event propagation status of every user, where $\br_t$'s $i$-th entry $[\br_t]_i=1$ if the event has ever been posted or retweeted by the $i$-th user since the beginning to the $t$-th time frame, and $[\br_t]_i=0$ otherwise. In addition, let $\br_0$ be a binary vector such that its nonzero entries indicate the set of users who first post the event. Then the event propagation model can be written as an iterative state equation 
	\begin{align}
	\label{eqn_retweet_propagation_model}
	\br_{t+1}=\TB \lb \br_t + \TB \lb \bA_{t+1} \br_t \rb \rb, ~\forall~t=0,1,2,\ldots,F-1,
	\end{align}
	where $\TB(\cdot)$ is an entry-wise threshold function defined as $[\TB(\bx)]_i=1$ if $[\bx]_i>1$ and $[\TB(\bx)]_i=[\bx]_i$ if $0 \leq [\bx]_i \leq 1$, for any nonnegative vector $\bx$. 
	The term  $\TB \lb \bA_{t+1} \br_t \rb$ can be viewed as the increment vector for event propagation in the $t+1$-th time frame.
	
	The derivation of the event propagation model in (\ref{eqn_retweet_propagation_model}) is as follows. Since $\bA_t$ accounts for the adjacency matrix of activated follower links for event propagation during the $t$-th time frame, the $i$-th entry of the vector $\bA_{t+1} \br_t$ can be expressed as $[\bA_{t+1} \br_t]_i=\sum_{j=1}^n [\bA_{t+1}]_{ij} [\br_t]_j$, which is the number of tweets involving the event that user $i$ decides to share on Twitter during the $t+1$-th time frame. Therefore, the entry-wise thresholded binary vector $\TB(\bA_{t+1} \br_t)$  indicates the status of new users participating in event propagation during the $t+1$-th time frame. Lastly, since $\TB(\bA_{t+1} \br_t)$ represents the vector of event propagation increment, $\br_{t+1}=\TB(\br_t+\TB(\bA_{t+1} \br_t))$ accounts for the event propagation status of all users since the beginning to the $t+1$-th time frame.
	
	The event propagation model in (\ref{eqn_retweet_propagation_model}) can be easily adapted to incorporate the NoN structure of a Twitter follower network. Let $\bAbet$ and $\bAwit$ denote the adjacency matrix of the between-network and within-network follower links, respectively. The event propagation model can be rewritten as 
	\begin{align}
	\label{eqn_retweet_propagation_model_NoN}
	\br_{t+1}=\TB  \lb \br_t + \TB \lb {\bAbet_{t+1}} \br_t \rb + \TB \lb {\bAwit_{t+1}} \br_t \rb \rb
	\end{align}
	for all $t=0,1,2,\ldots,F-1$. The matrices $\bAbet_t$ and $\bAwit_t$ are defined similarly as $\bA_t$ such that $\bA_t=\bAbet_t+\bAwit_t$. The terms  $\TB \lb {\bAbet_{t+1}}  \br_t \rb$ and  $\TB \lb {\bAwit_{t+1}} \br_t \rb$ in (\ref{eqn_retweet_propagation_model_NoN}) account for the event propagation increment caused by between-network and within-network follower links, respectively.

	\subsection{Surrogate function for event propagation}
	\label{subsec_surrogate}
	
	We are interested in investigating the effect of link removals on a Twitter follower network prior to actual event propagation. Correspondingly, only the adjacency matrix $\bA$ of the Twitter follower network is known, whereas the event propagation status vector $\br_t$ and the adjacency matrix $\bA_t$ affecting actual event propagation are unknown. Nonetheless, we will show that the largest eigenvalue of $\bA$, denoted by $\lambda_{\max}(\bA)$, can be used as a surrogate function for the containment of event propagation, as it is associated with an upper bound on the increment of event propagation. 
	Interestingly, $\lambda_{\max}(\bA)$ is known to be related to the information dissemination threshold of some parametric epidemic models \cite{Pastor-Satorras15epidemic_review,prakash2012threshold}. For the purpose of analysis, we assume the matrix $\bA$ is irreducible. This is equivalent to the assumption that in the Twitter follower graph there exists a path between any pair of nodes. The same assumption is also used in analyzing the NetMelt algorithm in \cite{tong2012gelling}. If $\bA$ is not irreducible, our analysis can be applied to any irreducible submatrix of $\bA$. 
		For ease of analysis, we assume the considered adjacency matrices (e.g., $\bA$ and $\bA_{t}$) are diagonalizable. If they are not diagonalizable, one could resort to spectral projections onto the Jordan subspaces of the adjacency matrix \cite{Deri_spectral17}.

	Specifically, let $\|\bx\|_0$ denote the number of nonzero entries of an $n$-dimensional vector $\bx$, which is also known as the $\ell_0$ norm or the sparsity level of $\bx$. Under the sparsity assumption that $\| \br_F \|_0 \leq s$, where $s \leq n$ is a trivial upper bound on $s$, we can obtain a surrogate function of the increment $\|\TB \lb \bA_{t+1} \br_t  \rb\|_0$ in terms of  $\lambda_{\max}(\bA)$ and $s$, which is
	\begin{align}
	\label{eqn_surrogate_retweet}
\left\|\TB \lb \bA_{t+1} \br_t  \rb \right\|_0 \leq 
	 C \cdot \sqrt{s} \cdot \lambda_{\max}(\bA)
	\end{align} 
for all $t=0,1,2,\ldots,F-1$, where $C$ is some constant that depends only on $\{\bA_{t+1}\}_{t=0}^{F-1}$. In other words, given $\{\bA_{t+1}\}_{t=0}^{F-1}$, the constant $C$ is independent of $\bA$.
 The derivation is given in Appendix \ref{appen_UB}.
	Since we consider the practical scenario where only the adjacency matrix $\bA$ is given, while the source of the event propagation and the values of $s$ and $\{\bA_t\}$ are unknown, $\lambda_{\max}(\bA)$ in (\ref{eqn_surrogate_retweet}) 
	serves as a proxy for the effect of the network topology on event propagation. 
%	In addition,  $\lambda_{\max}(\bA)$ is known to be associated with the information dissemination threshold of some parametric epidemic models \cite{Pastor-Satorras15epidemic_review,prakash2012threshold}. 			
	 It is clear from (\ref{eqn_surrogate_retweet}) that minimizing the largest eigenvalue $\lambda_{\max}(\bA)$ of the adjacency matrix $\bA$ can be effective in containing event propagation, since $\lambda_{\max}(\bA)$ is associated with an upper bound on the event propagation increment $\|\TB \lb \bA_{t+1} \br_t  \rb\|_0$  for each iteration in $t$.

	Applying the results in (\ref{eqn_surrogate_retweet}) to the event propagation model with NoN structure in (\ref{eqn_retweet_propagation_model_NoN}), we can obtain upper bounds on the increments $\TB \lb {\bAbet_{t+1}}  \br_t \rb$ and  $\TB \lb {\bAwit_{t+1}} \br_t \rb$ associated with between-network and within-network follower links in terms of $\lambda_{\max}(\bAbet)$ and $\lambda_{\max}(\bAwit)$, which are
	\begin{align}
	\label{eqn_surrogate_retweet_NoN}
	&\left\| \TB \lb {\bAbet_{t+1}} \br_t  \rb \right\|_0 \leq  
	C_{\textnormal{bet}} \cdot \sqrt{s}  \cdot \lambda_{\max}(\bAbet);  \\
	\label{eqn_surrogate_retweet_NoN_2}
	&\left\| \TB \lb {\bAwit_{t+1}} \br_t  \rb \right\|_0 \leq
		C_{\textnormal{wit}} \cdot \sqrt{s} \cdot  \lambda_{\max}(\bAwit),
	\end{align}
	where $C_{\textnormal{bet}}$ and $C_{\textnormal{wit}}$ are some constants, respectively.

	\subsection{LES: left eigenvector score}
	\label{subsec_LES}

	Since  $\lambda_{\max}(\bA)$ is proxy for event propagation, 
in what follows we first consider a generic form of score functions defined on follower links that ranks those links that would maximize reduction of the largest eigenvalue, and then propose to use the leading left eigenvector $\by$ of the adjacency matrix $\bA$ to define a score for each follower link that captures its importance in event propagation.

	Let $(i,j)$ denote a follower link in the Twitter follower network representing the relation that user $i$ follows user $j$.
	We consider the follower link score for assessing the influence in event propagation taking the form
\begin{align}
\label{eqn_edge_score}
\textnormal{score}(i,j)=[\bx]_i \cdot [\bx]_j,
\end{align}
where $\bx$ is an $n$-dimensional nonnegative vector with unit length. The following analysis shows the effect of link removals based on (\ref{eqn_edge_score}) on reducing the largest eigenvalue. 	Let $\cE_{\cR}$ denote a subset of follower links in a Twitter follower network such that
$(i,j) \in \cE_{\cR}$ if the follower link $(i,j)$ will be removed from the Twitter follower network. 
For any follower link removal set $\cE_{\cR}$ with cardinality $|\cE_{\cR}|=q \geq 1$,  let $\bAt(\cE_{\cR})$ be the adjacency matrix after removing the follower links in $\cE_{\cR}$ from the Twitter follower network. Under the assumption that $\bAt(\cE_{\cR})$ is diagonalizable, we can obtain upper and lower bounds on $ \lambda_{\max}(\bAt(\cE_{\cR}))$ in terms of $\lambda_{\max}(\bA)$ and $\bx$ as follows:
	%	If $\sum_{(i,j)\in \cE_{\cR}} [\by]_i [\by]_j >0$, then
	\begin{align}
	\label{eqn_event_propagation_bound_general}	
	\lambda_{\max}(\bA) - c \cdot \sum_{(i,j)\in \cE_{\cR}} [\bx]_i [\bx]_j \geq \lambda_{\max}(\bAt(\cE_{\cR})),
	\end{align}
	where $c$ is some positive constant that depends on the choice of $\bx$.
The proof is given in Appendix \ref{appen_UB_2}.

In this paper, we propose to use the leading left eigenvector $\by$ of $\bA$ to compute the link score, which we call the Left Eigenvector Score (LES).  The LES is  defined as
\begin{align}
\label{eqn_LES}
\textnormal{LES}(i,j)=[\by]_i \cdot [\by]_j.
\end{align}
Since $\by$ is the vector of eigenvector centrality, the LES of a link $(i,j)$ is the product of the associated eigenvector centrality $[\by]_i \cdot [\by]_j$. As a result,
high LES for a follower link $(i,j)$ means that the followers of both user $i$ and user $j$ play an important role in the Twitter follower network, and hence the follower link $(i,j)$ is crucial to event propagation.

%can show that removing the follower links of top LES can be effective in reducing the resulting surrogate function, and hence is able to contain event propagation increment according to (\ref{eqn_surrogate_retweet}).	

 By the Perron-Frobenius theorem \cite{HornMatrixAnalysis}, the largest eigenvalue of an adjacency matrix is always real and nonnegative, and its associated left eigenvector $\by$ has nonnegative entries and unit Euclidean norm, i.e., $[\by]_i \geq 0$ for all $i$ and $(\sum_i [\by]_i^2)^{1/2}=1$. Since $\by$ satisfies the relation $\bA^T \by=\lambda_{\max}(\bA) \by$, where $\cdot^T$ denotes the transpose of a matrix (or a vector), $\by$ is the vector of eigenvector centrality of each user based on every user's follower connectivity pattern in the Twitter follower network \cite{Newman10NetworkIntro}. Specifically, since the $i$-th entry of $\by$ has the relation $[\by]_i=\frac{1}{\lambda_{\max}(\bA)}\sum_{j=1}^n \bA_{ji} \by_j$, a user's importance is proportional to the sum of importance of his/her followers.

Moreover, when LES is used to reduce the largest eigenvalue via link removals, we can obtain both upper and lower bounds on $ \lambda_{\max}(\bAt(\cE_{\cR}))$ in terms of $\lambda_{\max}(\bA)$ and $\by$ as follows:
%	If $\sum_{(i,j)\in \cE_{\cR}} [\by]_i [\by]_j >0$, then
	\begin{align}
	\label{eqn_event_propagation_bound}
	\lambda_{\max}(\bA) - \sum_{(i,j)\in \cE_{\cR}} [\by]_i [\by]_j \leq \lambda_{\max}(\bAt(\cE_{\cR})); \\  
	\label{eqn_event_propagation_bound_2}	
	\lambda_{\max}(\bA) - c \cdot \sum_{(i,j)\in \cE_{\cR}} [\by]_i [\by]_j \geq \lambda_{\max}(\bAt(\cE_{\cR})),
	\end{align}
	where $c$ is some positive constant.
The proof is given in Appendix \ref{appen_UB_2}.
The bounds in (\ref{eqn_event_propagation_bound}) and (\ref{eqn_event_propagation_bound_2}) quantify the effect of LES-based link removals on the resulting largest eigenvalue $\lambda_{\max}(\bAt(\cE_{\cR}))$.	
	 We note that the bound in (\ref{eqn_event_propagation_bound}) is specific to the leading left eigenvector $\by$, whereas the bound in (\ref{eqn_event_propagation_bound_2}) can be easily obtained by setting $\bx=\by$ in (\ref{eqn_event_propagation_bound_general}).
Specifically, the bound in (\ref{eqn_event_propagation_bound_general}) motivates removing different top-score links for event propagation analysis, which will be discussed in Sec. \ref{sec_experiment}. These experimental results show that among all the compared link scores, the proposed LES is the most effective score in containing event propagation on Twitter, as $\by$ is the vector of eigenvector centrality of each node based on the follower link connectivity pattern.

	Similar analysis to (\ref{eqn_event_propagation_bound}) and (\ref{eqn_event_propagation_bound_2}) can be directly applied to the largest eigenvalues $\lambda_{\max}(\bAbet)$ and $\lambda_{\max}(\bAwit)$ in (\ref{eqn_surrogate_retweet_NoN}) and (\ref{eqn_surrogate_retweet_NoN_2}) by using their corresponding leading left eigenvectors. As a result, the proposed LES can be easily adapted to the NoN structure in the Twitter follower network.		

	\subsection{Computational complexity of LES-based link removal}
	Since the number of nonzero entries in $\bA$ is the total number of follower links $m$, computing the leading left eigenvector $\by$ takes $O(m)$ time by power iteration methods,  and reporting the top $q$ follower links of LES takes $O(mq)$ time. Therefore, the overall computational complexity for finding the removal set $\cE_{\cR}$ of cardinality $q$ is $O(mq)$. In principle, one is often interested in identifying a small set of influential links relative to the entire links. The parameter $q$ can either be a user-specified value that is application-dependent, or be determined by selecting the link removal set that maximizes the ratio of the corresponding sum of score function of the link removal set to the number of removed links.

	\begin{table*}[t]
		\centering
		\caption{Statistics of the collected events and Twitter follower networks}
		\label{Table_event_propagation}
		\begin{tabular}{c|c|c|c|c|c|c}
			
			\hline
			Dataset    & Event             & Users     & Follower Links & \begin{tabular}[c]{@{}c@{}}Networks \\ (Languages)\end{tabular} & \begin{tabular}[c]{@{}c@{}}Between-Network \\ Follower Links\end{tabular} & \begin{tabular}[c]{@{}c@{}}Within-Network \\ Follower Links\end{tabular} \\ \hline
			Obama FB   & http://Facebook.com/POTUS & 5,169,477 & 7,272,858      & 117                                                             & 19.74\%                                                                   & 80.26\%                                                                  \\
			Premier 12 & \#premier12               & 7,557,534 & 9,702,942      & 90                                                              & 22.11\%                                                                   & 77.89\%                                                                  \\
			AlphaGo    & \#AlphaGo                 & 9,259,187 & 9,794,702      & 141                                                             & 29.35\%                                                                   & 70.65\%                                                                  \\
			\hline
		\end{tabular}
	\end{table*}
	
	\section{Experiments on Synthetic Datasets and Twitter Traces}
	\label{sec_experiment}
	\subsection{Experiment setup and dataset description}
	\label{subsec_experiment_setup_NoN}
To study the effect of follower link removals on event propagation, in this section we conducted two types of experiments: (I) simulated event propagation on synthetic random graphs and (II) real-world event propagation traces collected from Twitter API.

		 For (I), we generated $1000$ synthetic random graphs consisting of two groups of non-overlapping users with the following follower link generation parameters. The number of users in group $g$ is denoted by $n_g$, $g=1,2$. For any two users $i$ and $j$ in the network, $i \neq j$, node $i$ follows node $j$ with probability $p_{g_i g_j}$, where $g_i,g_j \in \{1,2\}$ are the group labels of $i$ and $j$, respectively.
		To simulate the traces of event propagation, we randomly select $n_{\textnormal{ini}}$ nodes from group 1 as the source and implement the independent cascade model (ICM) \cite{goldenberg2001using} for event propagation. Given the $n_{\textnormal{ini}}$ initially activated nodes and the generated follower network,
		the ICM iteratively activates new follower links for event propagation via the follower network topology. Here each follower link can be activated  based on certain activation probability and  at most once, and
		we use the trivalency model \cite{chen2010scalable} to assign the activation probability, which is uniformly selected from the set $\{1,0.1,0.01\}$. For performance evaluation, we designed two sets of parameters with identical group sizes and average number of follower links.		
		 The first dataset (Dataset 1) emphasizes the importance of between-network links for event propagation, as there are fewer between-network links than within-network links, and all between-network links are responsible for event propagation from group 1 to group 2 by setting $p_{12}=0$ and $p_{11}=p_{22}>p_{21}$.  The second dataset (Dataset 2) emphasizes the importance of within-network links for event propagation, as the event propagation only occurs within group 1 when we set the parameter $p_{21}=0$.

	For (II), we collected three real-world event propagation traces and user languages from Twitter as described in Sec. \ref{sec_NoN_structure}.
	We also collected the users who have seen but have not retweeted the event (i.e., non-retweeters) 
	and their user language settings to form a Twitter follower  network and a NoN model for testing the effect of link removals on event propagation, where we assume following entails viewing tweets. 
	In other words, the collected Twitter follower networks include the follower connectivity structure of retweeters and non-retweeters of an event, and their user languages are used to identify the NoN structure. 
	The statistics of the collected datasets are summarized in Table \ref{Table_event_propagation}. One notable NoN feature of these Twitter follower  networks is that the between-network follower links only account for a portion of from 20\% to 30\% of the total number of follower links.

	\textbf{Evaluation Metrics.} A link score function for assessing link influence on event propagation is a function of the adjacency matrix and the NoN model of a Twitter follower network. 	
	The actual event propagation traces are only used to compare the performance of different link scores. We use the \textit{event reachability} as the performance metric, which is defined as the fraction of users who can still post or retweet the events after some follower links are removed from the original Twitter follower network. The event fails to propagate further to a user's follower if the corresponding follower link has been removed. As a result, the set of link removals	
	 that lead to lower event reachability are the links that have more influence on event propagation.
	 
	 Specifically, for experiment (I), the event reachability $R_f(\cE_\cR)$ of a score function $f$ is defined as the average number of activated users with respect to a link removal set $\cE_{\cR}$ from $f$. For experiment (II), given the real-world event propagation traces, let $n_0$ denote the total number of users who have posted/retweeted the event. The event reachability $R_f(\cE_\cR)$  of a score function $f$ subject to a link removal set $\cE_{\cR}$ is defined as the number of remaining active retweeters divided by $n_0$.
	 	In addition, we use the average reachability $R^q_{\textnormal{rand}}$ of $q$ random link removals as the baseline performance, and define the efficiency of a score function $f$ as 
	 	\begin{align}
	 	\label{eqn_efficiency}
	 	\textnormal{efficiency}(\cE_{\cR})=\frac{R^{|\cE_{\cR}|}_{\textnormal{rand}}-R_f(\cE_\cR)}{R^{|\cE_{\cR}|}_{\textnormal{rand}}},
	 	\end{align}
	 	where $|\cE_{\cR}|$ denotes the size of the link removal set $\cE_{\cR}$, and higher efficiency means better performance in identifying influential links given the same number  of link removals. Throughout this paper, the results of random link removals are averaged over 10 trials.

	\textbf{Follower Link Scores.} We compare the effect of removing top $q$ follower links on event reachability based on different link score functions, for which the score function of a follower link ($i,j$) takes the form 
	\begin{align}
	\label{eqn_score_exp}
	\textnormal{score}(i,j)=[\bx]_i \cdot [\bxt]_j,
	\end{align}
	where $\bx$ and $\bxt$ are nonnegative $n$-dimensional vectors.
	The score function can be easily incorporated with centrality measures on users based on the Twitter follower network topology. However, score functions requiring the information of shortest paths among all node pairs, such as the edge betweenness \cite{Girvan02}, can be computationally demanding for large graphs. For instance, the Johnson's shortest algorithm \cite{johnson1977efficient} has computational complexity $O(nm+n^2 \log n)$, where $n$ and $m$ are the total number of users and follower links, respectively.
%	However, since the Twitter follower network is often not a connected graph, i.e., there is not a path connecting any two users in the network, centrality measures defined on connected graphs, such as the closeness and betweenness centrality measures \cite{Newman10NetworkIntro}, cannot be used as a score function.

	The following summarizes different score functions for performance comparison, including the scenario where the network identity of every user is known and the NoN model is applied such that the between-network and within-network follower links are used separately for link score computation. 
	The computational complexity of returning top $q$ follower links for different follower link score functions is summarized in Table \ref{table_complexity}. 	The implementation details and computational complexity analysis are given in Appendix \ref{appen_implementation}.

	\begin{figure*}[]
		\centering
		\begin{subfigure}[b]{0.4\linewidth}
			\includegraphics[width=\textwidth]{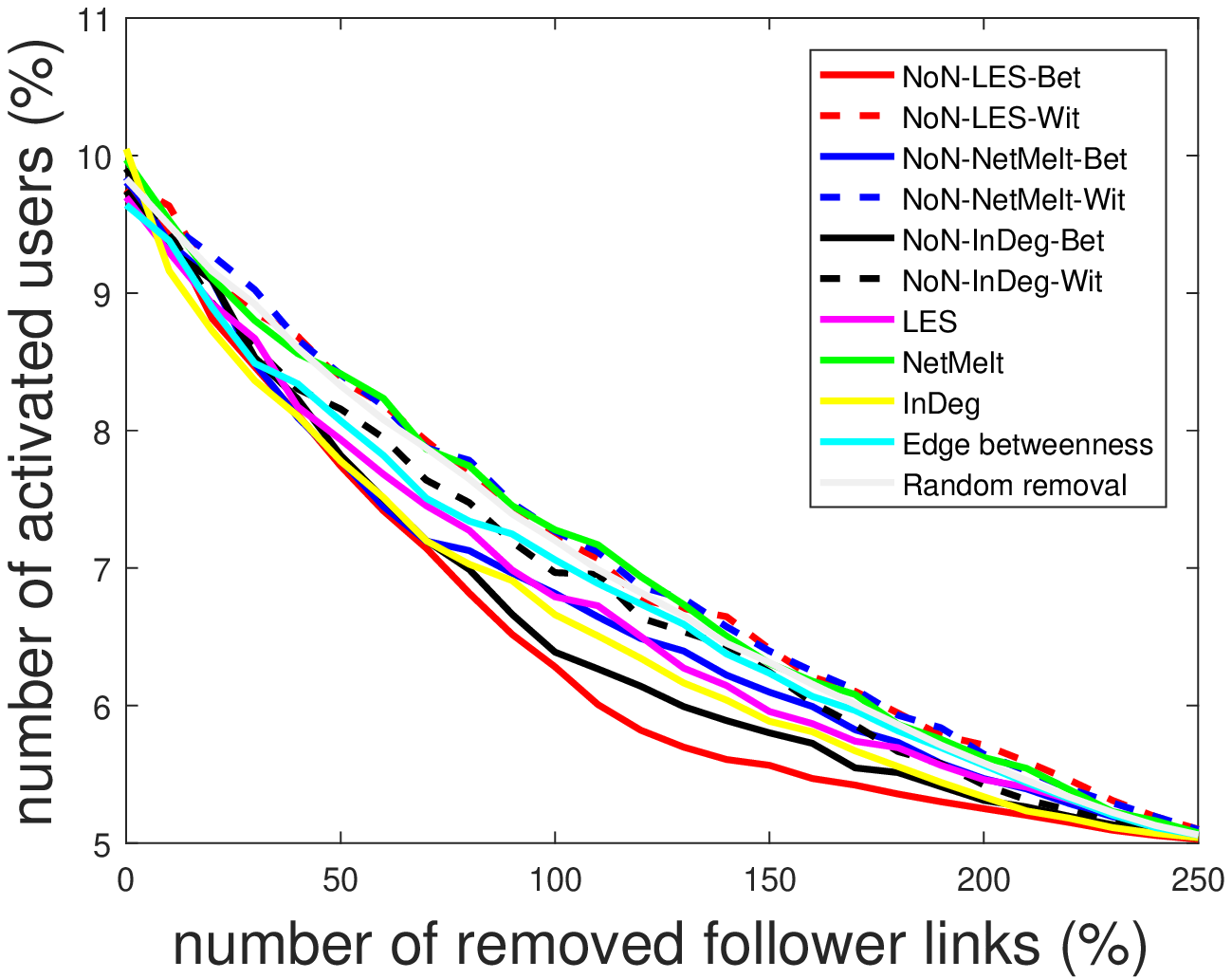}
			\caption{Dataset 1 - reachability}
			%			\label{}
		\end{subfigure}%
		%		\\
		%		\hspace{3.8cm}
		\centering
		\begin{subfigure}[b]{0.4\linewidth}
			\includegraphics[width=\textwidth]{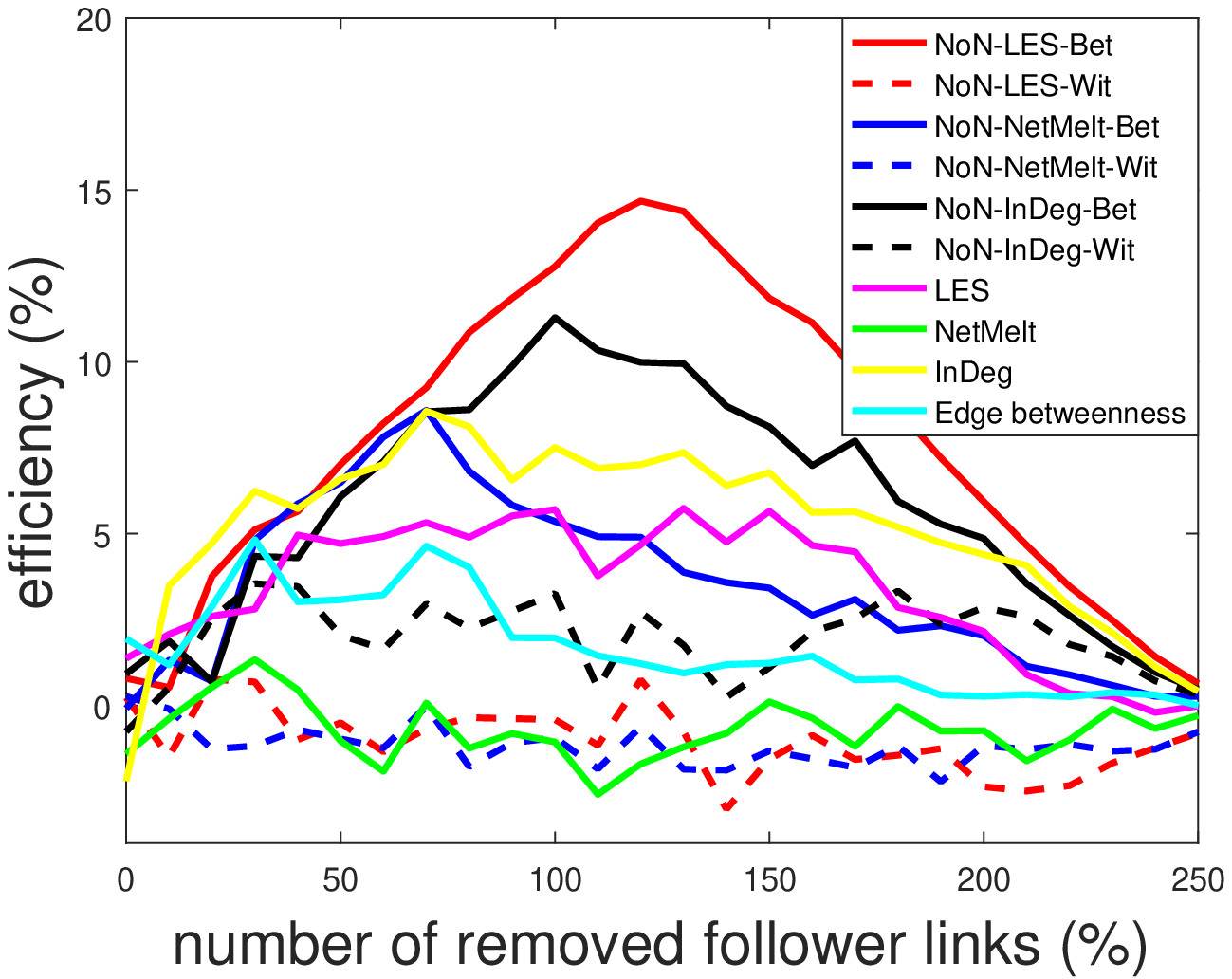}
			\caption{Dataset 1 - efficiency}
			%			\label{}
		\end{subfigure}
		%		\\
		\\
		\centering
		\begin{subfigure}[b]{0.4\linewidth}
			\includegraphics[width=\textwidth]{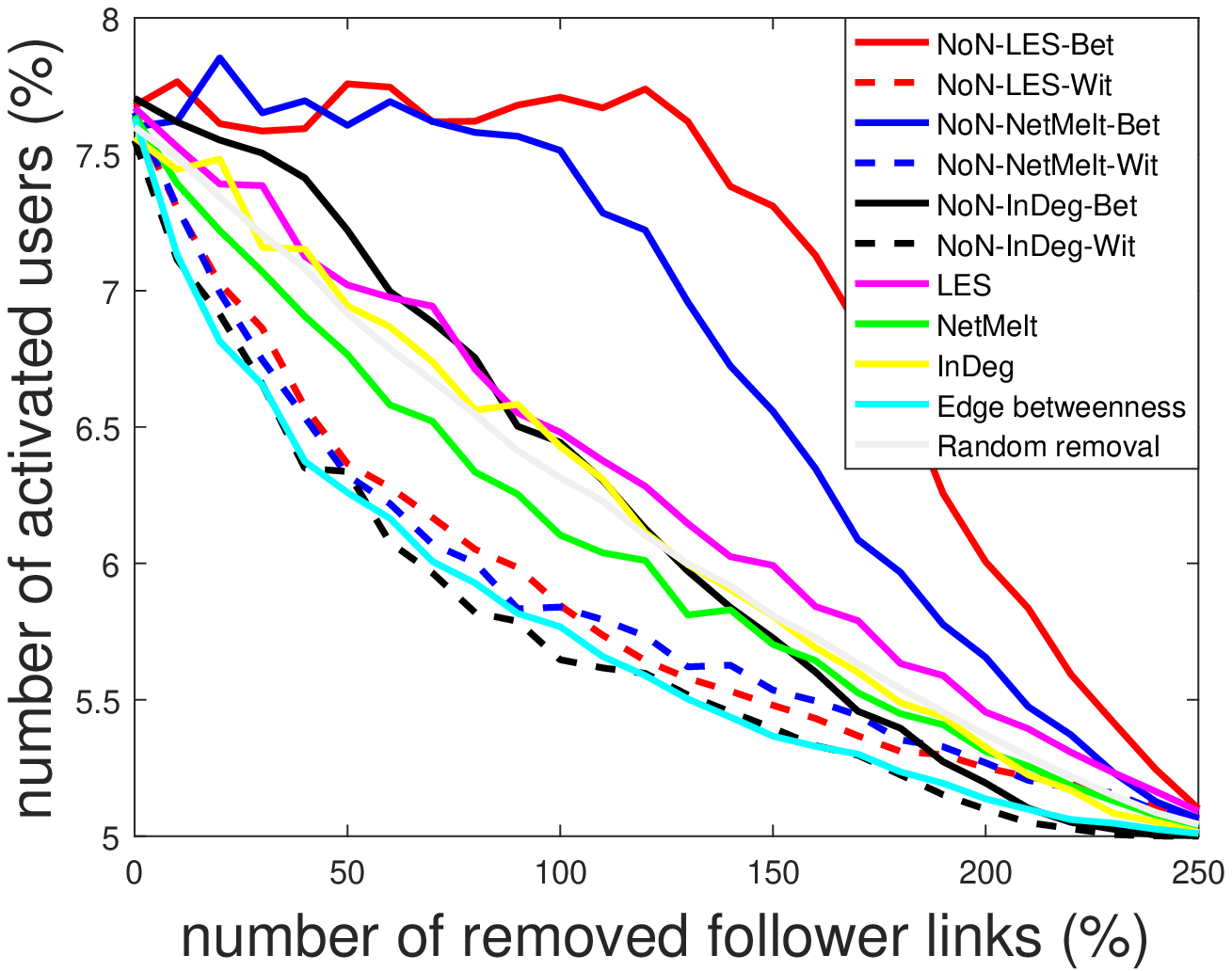}
			\caption{Dataset 2 - reachability}
			%			\label{}
			%			\label{}			
		\end{subfigure}				
		\centering
		\begin{subfigure}[b]{0.4\linewidth}
			\includegraphics[width=\textwidth]{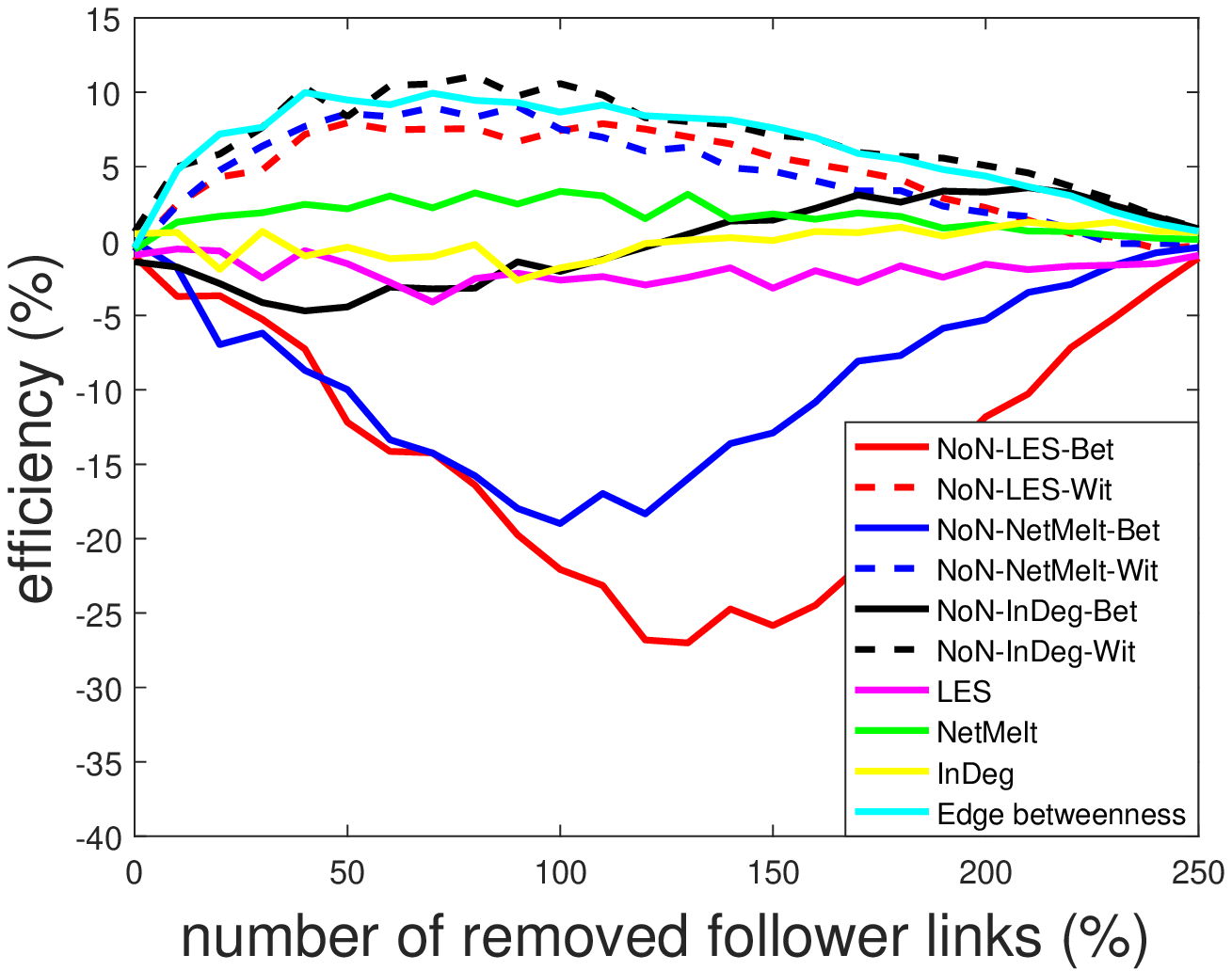}
			\caption{Dataset 2 - efficiency}
			%			\label{}			
		\end{subfigure}	
		\caption{
				The effect of removing top-score follower links on two synthetic event propagation datasets. 
				LES, NetMelt, InDeg and Edge-betweenness methods do not incorporate NoN information while the others use this information.  
				  Figure \ref{Fig_event_prop_sim} (a) and Figure \ref{Fig_event_prop_sim} (b) display the average reachability and efficiency of different score functions in the first dataset (Dataset 1) with parameters $n_{\textnormal{ini}}=5$, $n_1=n_2=100$, $p_{11}=p_{22}=0.01$, $p_{12}=0$, and $p_{21}=0.005$, which emphasizes the importance of between-network links for event propagation.  Figure \ref{Fig_event_prop_sim} (c) and Figure \ref{Fig_event_prop_sim} (d) display the average reachability and efficiency of different score functions  in the second dataset (Dataset 2) with parameters $n_{\textnormal{ini}}=5$, $n_1=n_2=100$, $p_{11}=p_{12}=0.01$, $p_{22}=0.005$, and $p_{21}=0$, which emphasizes the importance of within-network links for event propagation. The results show that  incorporating the NoN structure for computing score functions lead to improved performance for the synthetic datasets.				
			} 
			\label{Fig_event_prop_sim}
			%			\vspace*{-3mm}
		\end{figure*}

	\begin{itemize}
		\item \textbf{LES:} LES uses the leading left eigenvector of the adjacency matrix $\bA$ for score computation.
		
		\item \textbf{InDeg:} InDeg uses the in-degree (number of followers) of each user for score computation.
		
		\item \textbf{NetMelt:} NetMelt \cite{tong2012gelling} is an edge removal algorithm proposed to decrease the largest eigenvalue $\lambda_{\max}(\bA)$ via link removal by using the leading left and  right eigenvectors of $\bA$.		
			\item \textbf{Edge betweenness:} edge betweenness \cite{Girvan02} requires the information of shortest paths among all node pairs in a network. The importance of an edge is evaluated by	the number of shortest paths that pass through it.

		\item \textbf{NoN-LES-Bet (NoN-LES-Wit):} NoN-LES-Bet  (NoN-LES-Wit) 
		exploits the NoN structure and evaluates the score function using the leading left eigenvector of the between-network (within-network) adjacency matrix $\bAbet$ ($\bAwit$). 
		
		\item \textbf{NoN-InDeg-Bet (NoN-InDeg-Wit):} NoN-InDeg-Bet and NoN-InDeg-Wit are extensions of the InDeg score tailored to the NoN structure. 	
		
		\item \textbf{NoN-NetMelt-Bet (NoN-NetMelt-Wit):} Non-NetMelt-Bet and NoN-NetMelt-Wit are NetMelt algorithms that incorporate the NoN structure. 
	\end{itemize}

\begin{table}[!h]
	\centering
	\caption{Summary of computational complexity of returning top $q$ follower links for different follower link score functions. $m$ and $n$ are the total number of follower links and users, respectively.}
	\label{table_complexity}
	\begin{tabular}{|c|c|}
		\hline
		Score function                                                              & Complexity \\ \hline
		LES                                                                         & $O(mq)$    \\ \hline
		InDeg                                                                       & $O(mq)$    \\ \hline
		NetMelt                                                                     & $O(mq+n)$  \\ \hline
		Edge betweenness                                                                     & $O((n+q)m+n^2 \log n)$  \\ \hline	
		\begin{tabular}[c]{@{}c@{}}NoN-LES-Bet\\ (NoN-LES-Wit)\end{tabular}         & $O(mq)$    \\ \hline
		\begin{tabular}[c]{@{}c@{}}NoN-InDeg-Bet\\ (NoN-InDeg-Wit)\end{tabular}     & $O(mq)$    \\ \hline
		\begin{tabular}[c]{@{}c@{}}NoN-NetMelt-Bet\\ (NoN-NetMelt-Wit)\end{tabular} & $O(mq+n)$  \\ \hline
	\end{tabular}
\end{table}

	\begin{figure*}[]
		\centering
		\begin{subfigure}[b]{0.32\linewidth}
			\includegraphics[width=\textwidth]{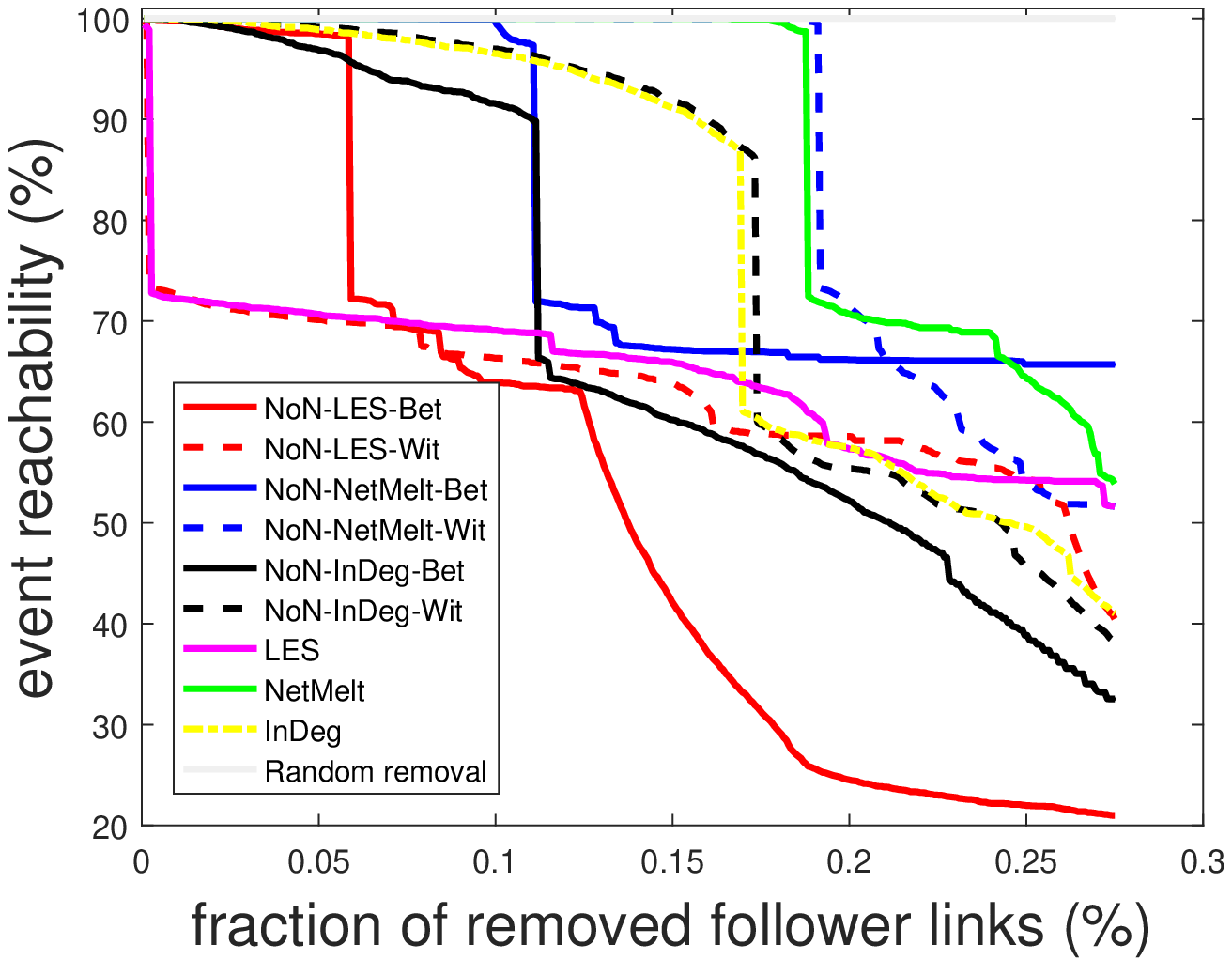}
			\caption{Obama FB - reachability}
			%			\label{}
		\end{subfigure}%
		\centering
		\begin{subfigure}[b]{0.32\linewidth}
			\includegraphics[width=\textwidth]{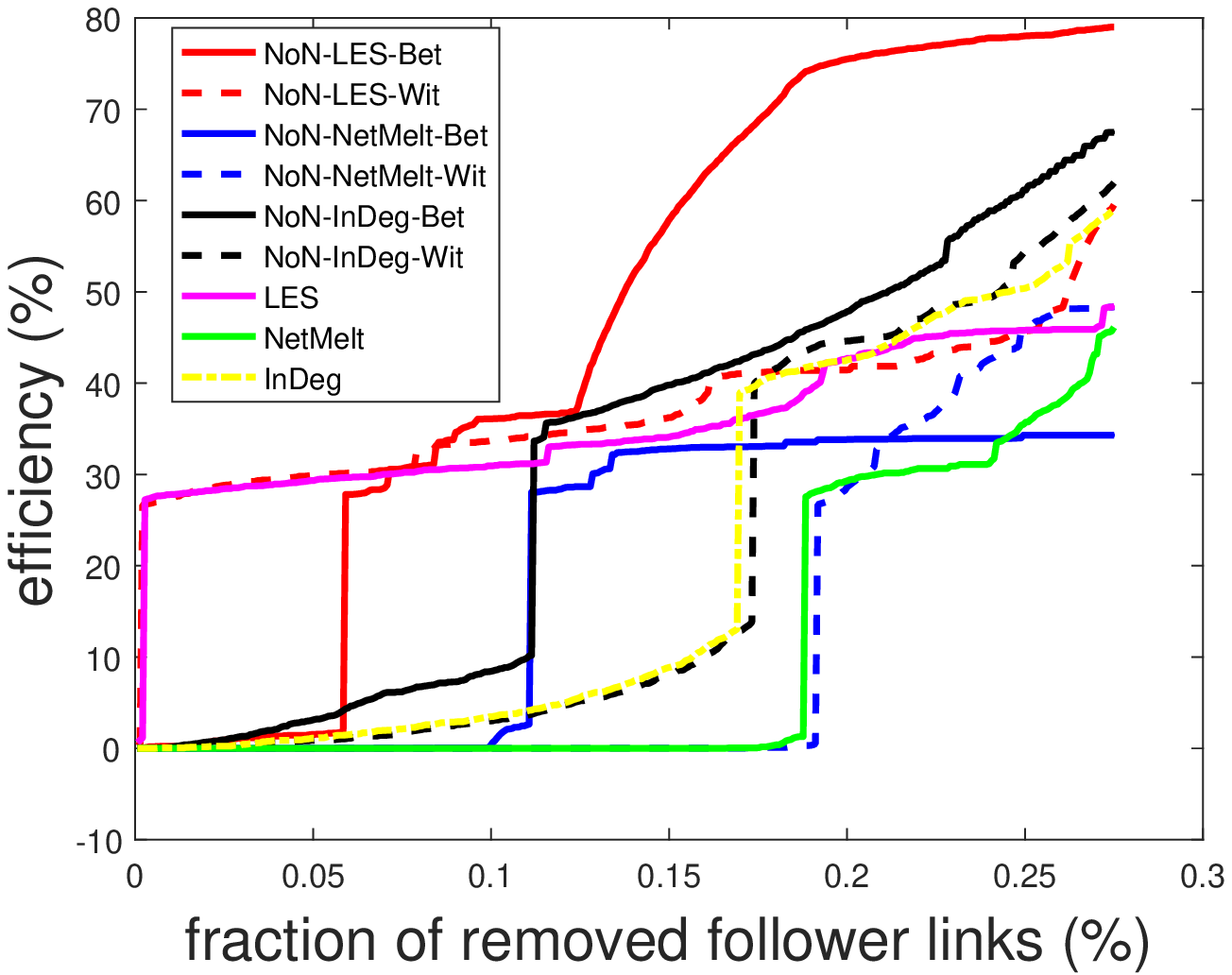}
			\caption{Obama FB - efficiency}
			%			\label{}
		\end{subfigure}%		
		%		\\
		%		\hspace{3.8cm}
		\centering
		\begin{subfigure}[b]{0.32\linewidth}
			\includegraphics[width=\textwidth]{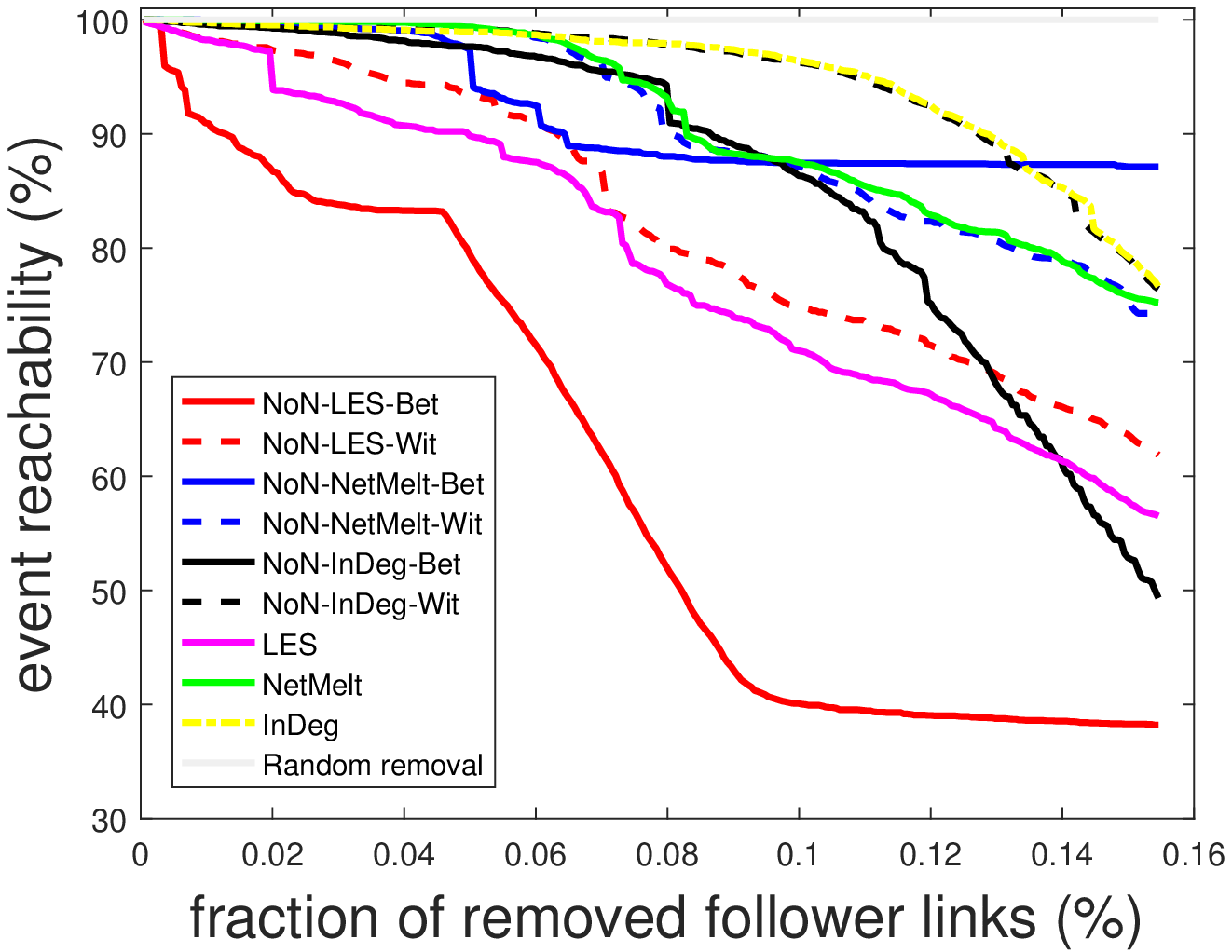}
			\caption{Premier 12 - reachability}
			%			\label{}
		\end{subfigure}
		\\
		\centering
		\begin{subfigure}[b]{0.32\linewidth}
			\includegraphics[width=\textwidth]{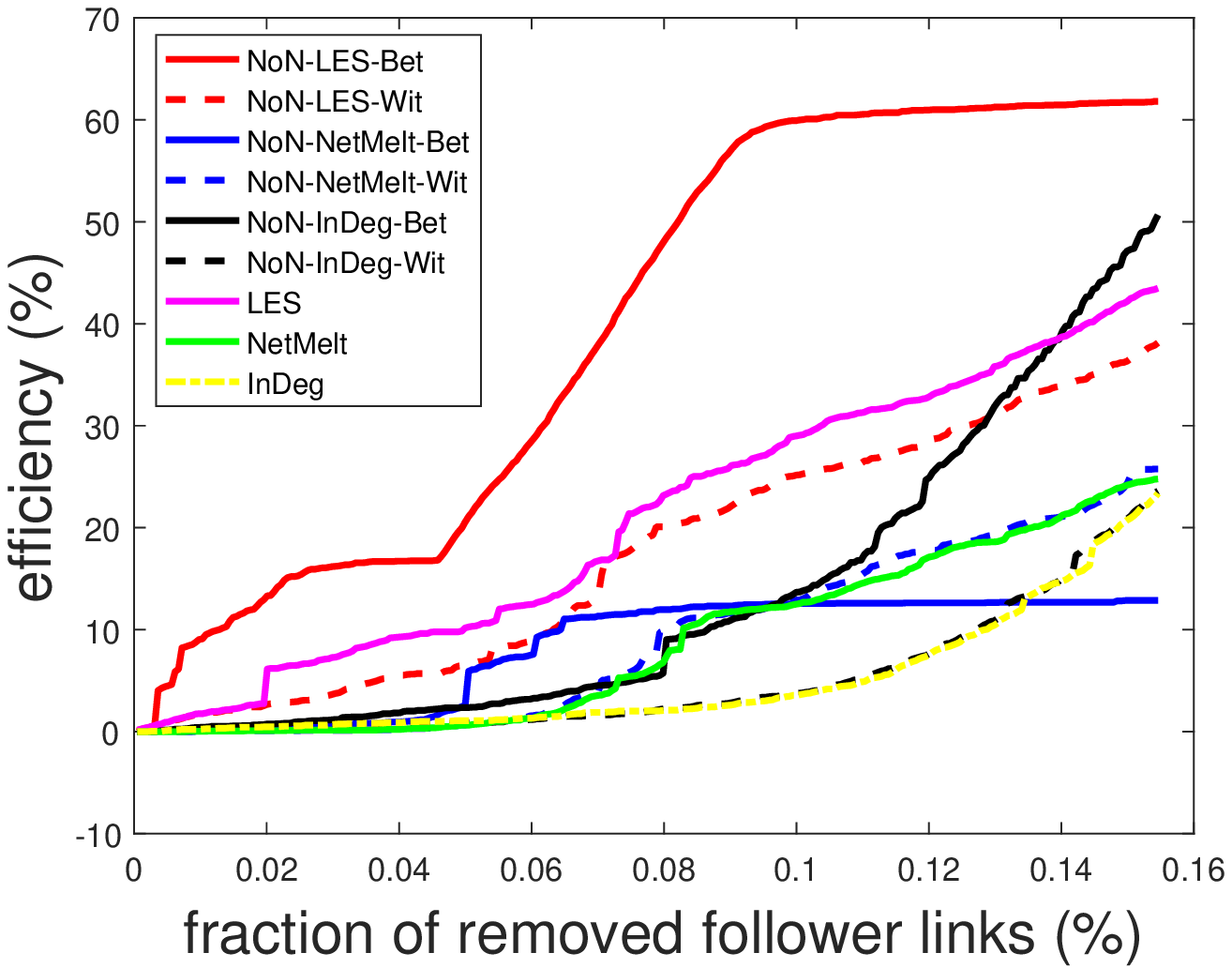}
			\caption{Premier 12 - efficiency}
			%			\label{}
		\end{subfigure}		
		%		\\
		\centering
		\begin{subfigure}[b]{0.32\linewidth}
			\includegraphics[width=\textwidth]{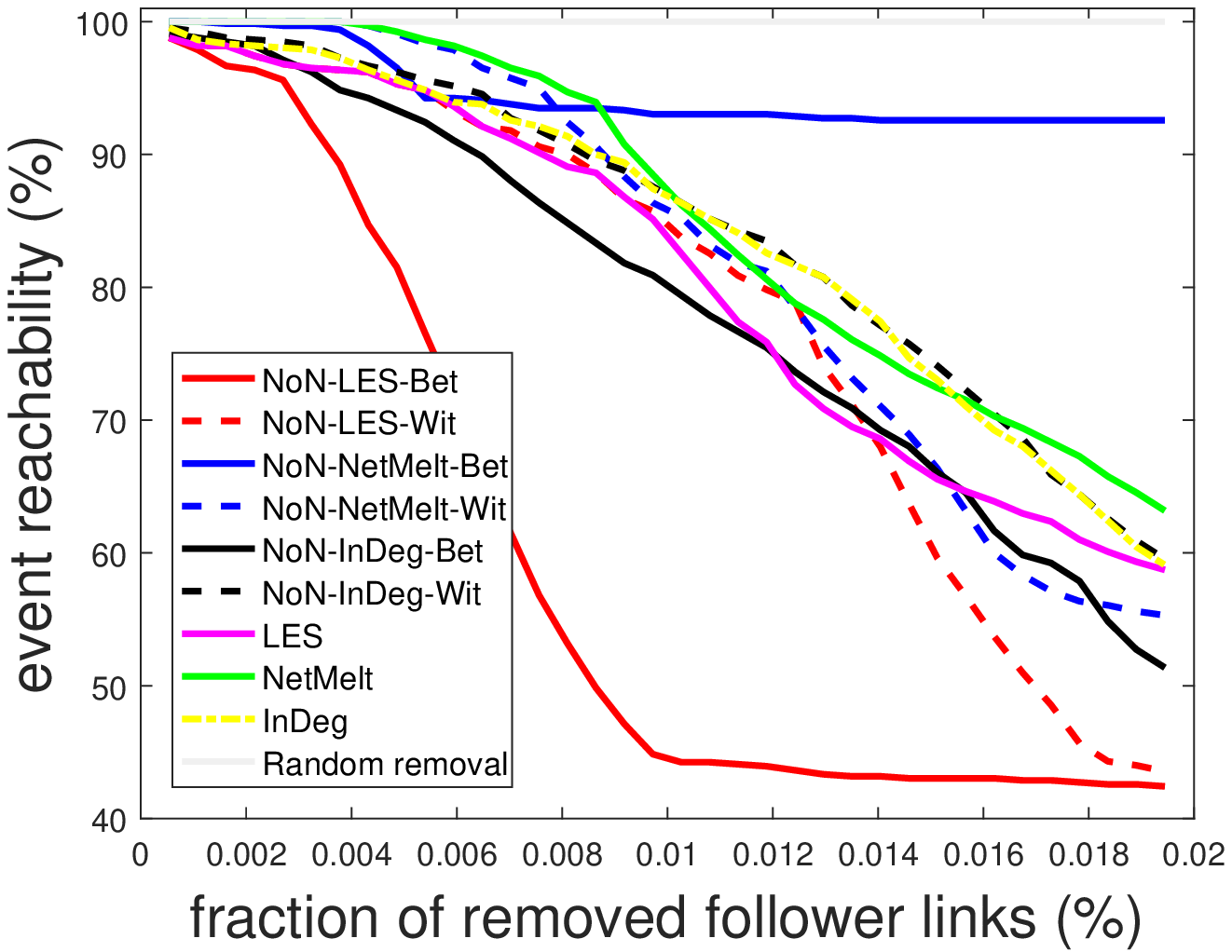}
			\caption{AlphaGo - reachability}
			%			\label{}
		\end{subfigure}					
		\centering
		\begin{subfigure}[b]{0.32\linewidth}
			\includegraphics[width=\textwidth]{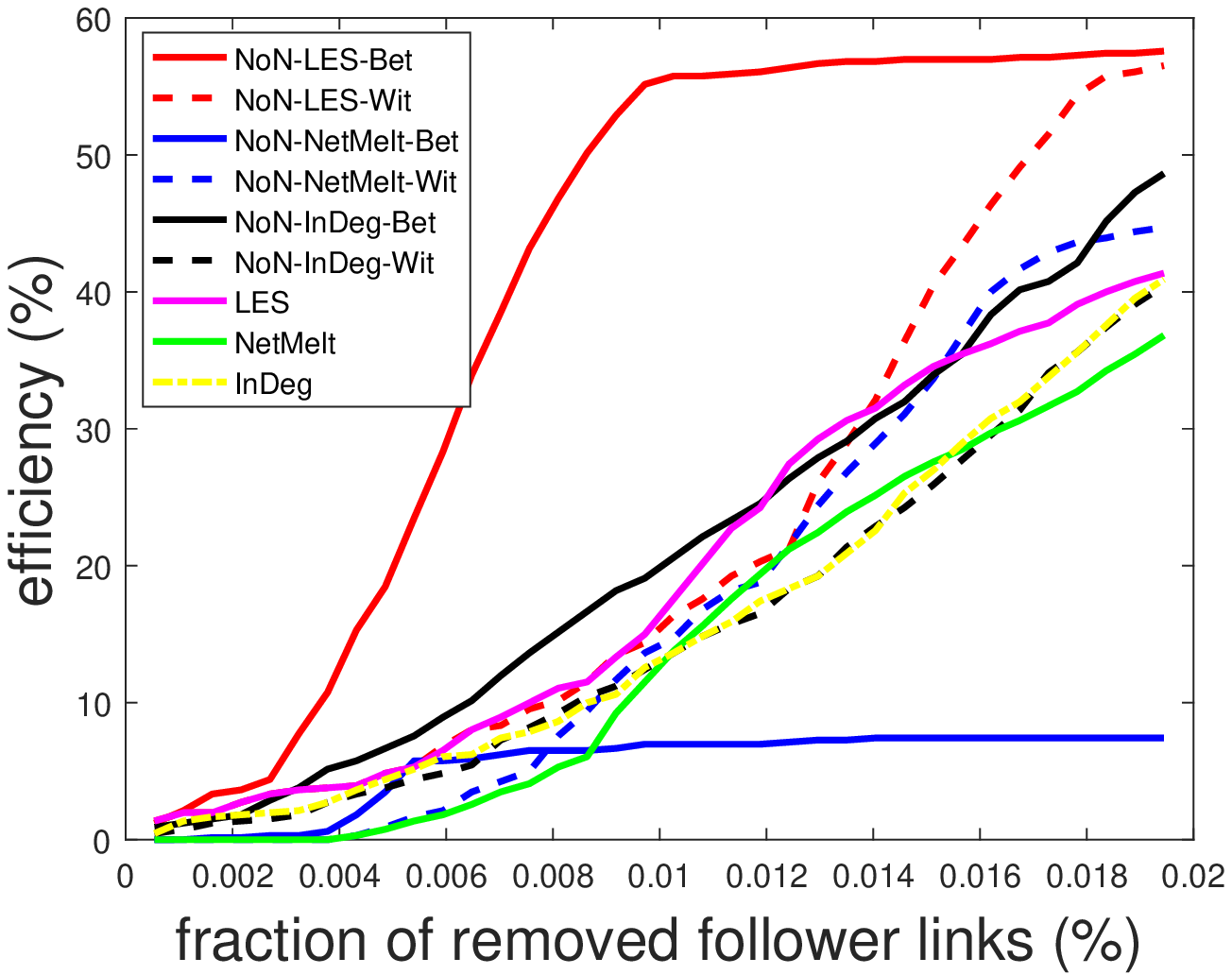}
			\caption{AlphaGo - efficiency}
			%			\label{}
		\end{subfigure}			
		\caption{The effect of removing top-score follower links on the collected Twitter datasets listed in Table \ref{Table_event_propagation}. Event reachability is the fraction of users who can still post or retweet the event after some follower links are removed from the original Twitter follower network. The efficiency of a score function is the performance improvement relative to random link removals, which is defined in (\ref{eqn_efficiency}).
			Observe that using the proposed LES and  exploiting the NoN structure  result in the highest efficiency (curve labeled NoN-LES-Bet).}
		\label{Fig_event_prop_control}
%			\vspace*{-3mm}
	\end{figure*}

	\subsection{Performance evaluation}
	Figure \ref{Fig_event_prop_sim} displays the event reachability and efficiency of event propagation in two synthetic event propagation datasets 
	with respect to  link removals based on different score functions as specified in Sec. \ref{subsec_experiment_setup_NoN}. The results in Figure \ref{Fig_event_prop_sim} show that incorporating the NoN structure for computing the score functions leads to better efficiency  than their counterparts without using the NoN structure. Specifically, NoN-LES-Bet, NoN-NetMelt-Bet and NoN-InDeg-Bet (NoN-LES-Wit, NoN-NetMelt-Wit and NoN-InDeg-Wit) have higher efficiency than LES, NetMelt and InDeg in the first (second) dataset, which demonstrates the advantage of using the NoN structure for identifying influential links. In the second dataset, edge betweenness has similar performance as NoN-LES-Wit, NoN-NetMelt-Wit and NoN-InDeg-Wit, whereas it has low efficiency in the first dataset.

	Figure \ref{Fig_event_prop_control} displays the event reachability and efficiency with respect to different link removal methods\footnote{We are unable to report the results of edge betweenness for the experiments on the collected Twitter follower networks, as edge betweenness is computationally more demanding than the other methods listed in Table \ref{table_complexity}.} as described in Sec. \ref{subsec_experiment_setup_NoN}. Comparing to the link removal methods without using the NoN structure (LES, InDeg  and NetMelt), it is observed that incorporating the NoN structure (user languages) can further reduce event reachability and result in better efficiency. In particular, the NoN-LES-Bet method outperforms other methods in the Premier 12 and AlphaGo datasets. For the Obama FB dataset, LES and NoN-LES-Wit can be more effective than other methods for the first few follower link removals. One possible explanation from Figure \ref{Fig_NoN_structure} is that in the Obama FB dataset the
	fraction of most-populated same-language retweeters (English retweeters) is more prominent than those in the other two datasets.  As a result, in the Obama FB dataset NoN-LES-Wit and LES can be more effective for the first few link removals. However, as the number of removals increases these two methods soon lose their appeals, and NoN-LES-Bet significantly outperforms other methods. For example, if we are able to remove 0.25\% of follower links from the Obama FB dataset, NoN-LES-Bet can reduce the event reachability to roughly 20\%, whereas the second best method (NoN-InDeg-Bet) only reduces the event reachability to roughly 35\%, which means that NoN-LES-Bet is 15\%  more effective in finding important links as compared to other methods. In these experiments, random link removals have limited effect on the reduction of event reachability, as the actual event propagation traces only involve a small subset of follower links.
	These results suggest that LES can better reflect the level of importance of a follower link for event propagation. More interestingly, the success of NoN-LES-Bet in reducing event propagation on Twitter implies that although between-network follower links correspond to under 30\% of the total number of follower links in these datasets, they are crucial to event propagation.

\begin{figure*}[]
	\centering
	\begin{subfigure}[b]{0.32\linewidth}
		\includegraphics[width=\textwidth]{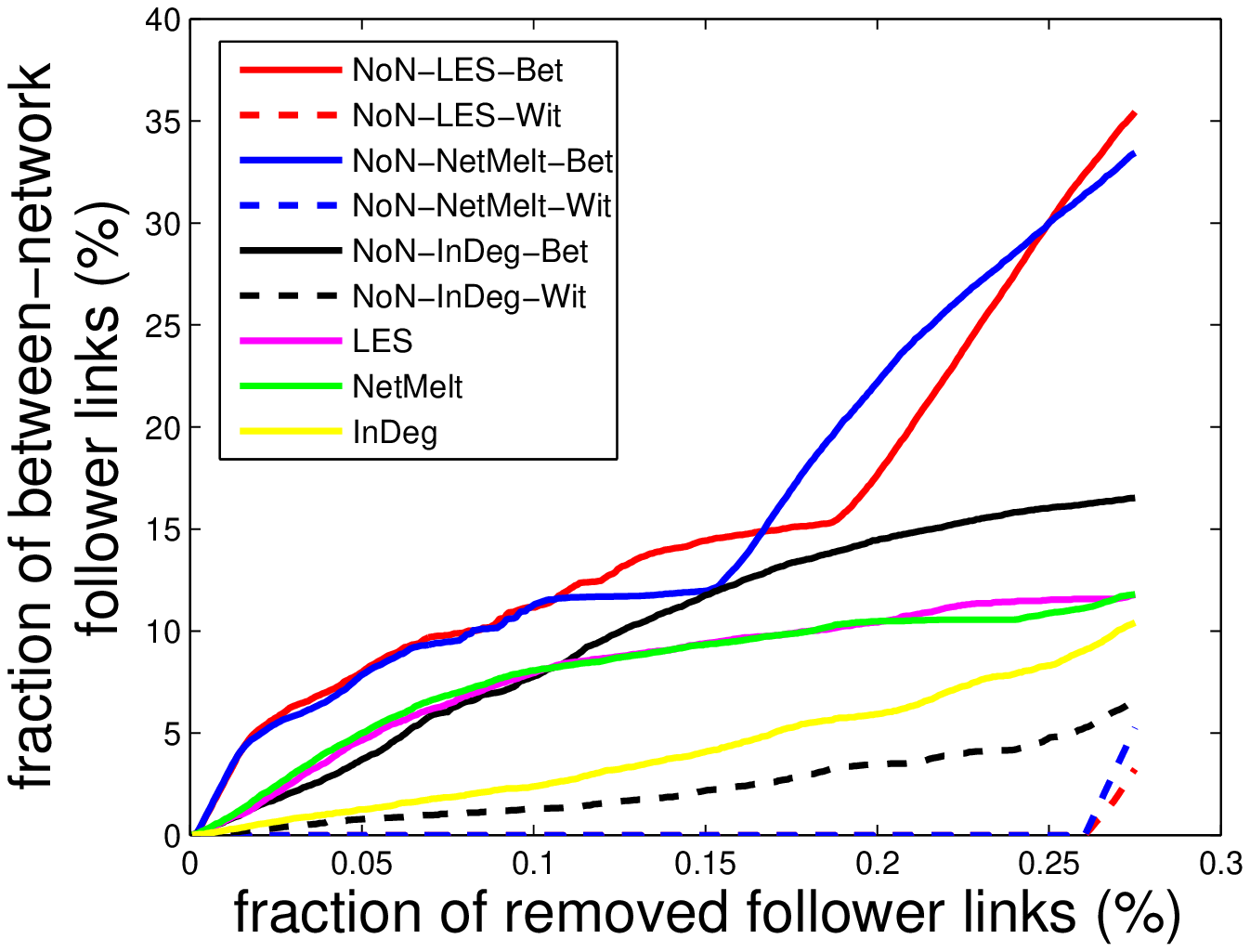}
		\caption{Obama FB}
		%			\label{}
	\end{subfigure}%
	%		\\
	%		\hspace{3.8cm}
	\centering
	\begin{subfigure}[b]{0.32\linewidth}
		\includegraphics[width=\textwidth]{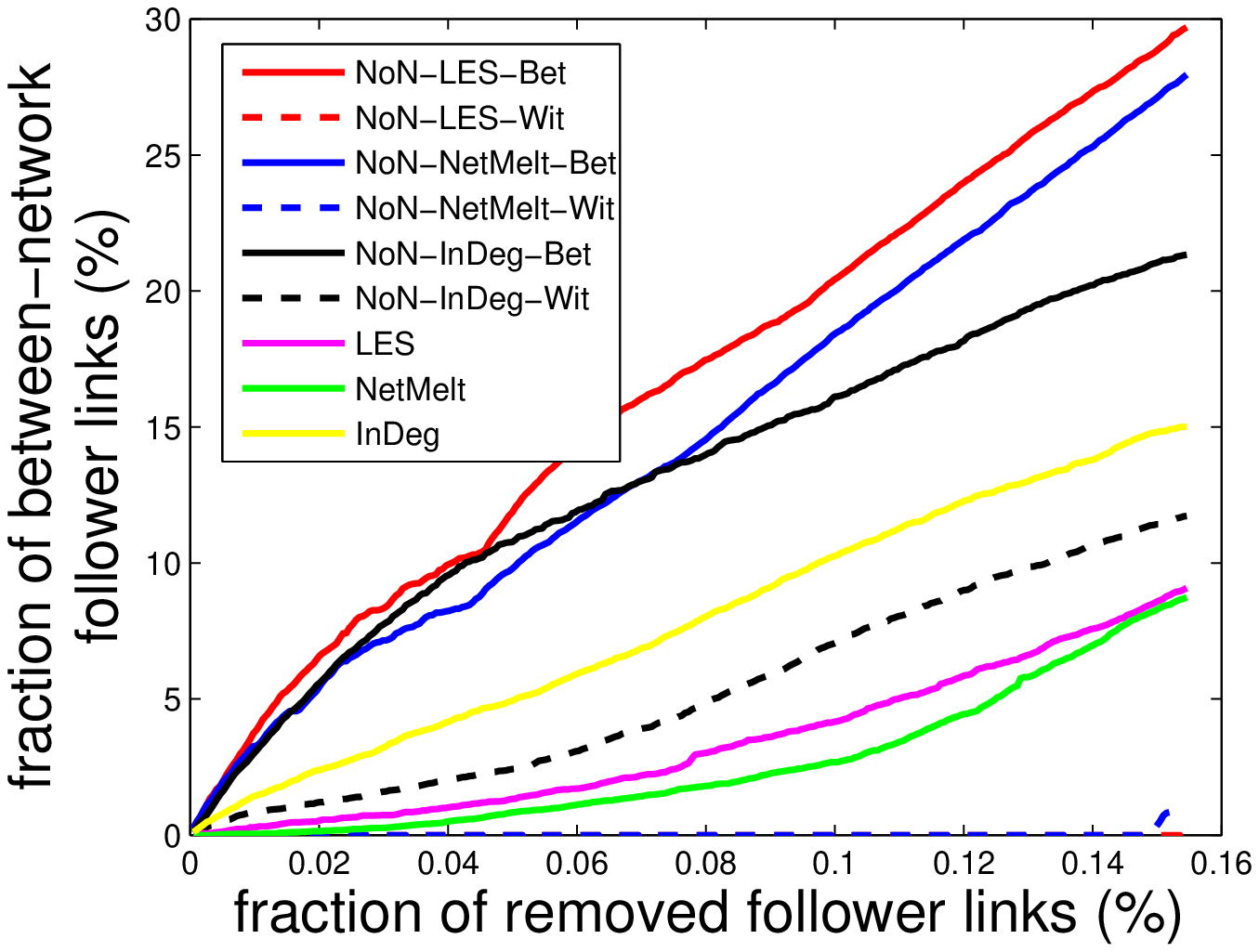}
		\caption{Premier 12}
		%			\label{}
	\end{subfigure}
	%			\\
	\centering
	\begin{subfigure}[b]{0.32\linewidth}
		\includegraphics[width=\textwidth]{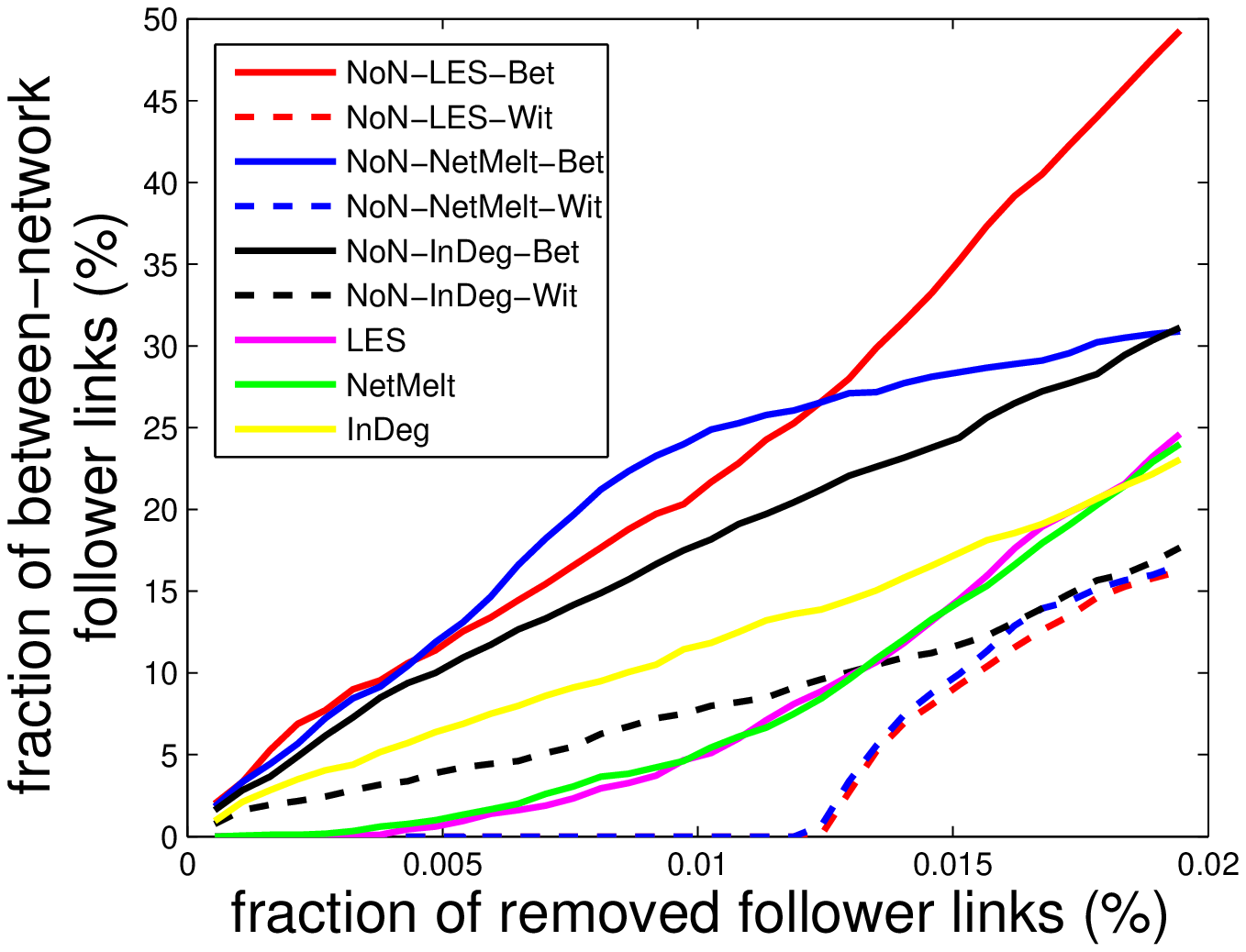}
		\caption{AlphaGo}
		%			\label{}
	\end{subfigure}		
	\caption{Fraction of between-network follower links  in the link removal set of different link removal methods. Comparing to Figure \ref{Fig_event_prop_control}, although the fraction of removed between-network follower links of NoN-LES-Bet and NoN-NetMelt-Bet are similar, the follower links identified by NoN-LES-Bet are more influential in event propagation as their removals result in lower event reachability.
	}
	\label{Fig_event_prop_rev}
	%	\vspace*{-2mm}
\end{figure*}

	The effectiveness of LES in reducing event propagation can be explained by the fact it is a minimizer of an upper bound on the increment of event propagation as established in Sec. \ref{sec_methodology_NoN}. In addition, as discussed in Sec. \ref{subsec_LES}, the leading left eigenvector $\by$ of the adjacency matrix $\bA$ that constitutes LES is the vector of eigenvector centrality of the follower network, where a user's importance is proportional to the sum of importance of his/her followers.	
	In contrast, link score functions based on in-degrees or NetMelt do not result in as much reduction as  compared with the LES-based methods.
	The finding that the LES-based methods are superior to the InDeg-based methods suggests that event propagation not only depends on the number of followers, but also on the role of each user's followers in event propagation. This is also consistent with the importance of social ties for event propagation in online social networks \cite{easley2010networks,tang2012inferring}. We also implemented score functions based on the leading right eigenvector of the adjacency matrix. However, its effect on reducing event propagation is not prominent, so we omit the results in the paper.

	Figure \ref{Fig_event_prop_rev} displays the fraction of between-network follower links in the link removal set of different link removal methods. 
It can be observed that the NoN model can indeed be used to emphasize between-network and within-network links. In particular, NoN-LES-Bet, NoN-NetMelt-Bet and NoN-InDeg-Bet lead to the selection of more between-network links in the corresponding link removal set when compared with LES, NetMelt and InDeg, respectively. Similarly, NoN-LES-Wit, NoN-NetMelt-Wit and NoN-InDeg-Wit favor the selection of  within-network links.
		In addition,
	we find that although NoN-LES-Bet and NoN-NetMelt-Bet lead to similar fraction of between-network follower link removals, 
	NoN-LES-Bet achieves lower event reachability than NoN-NetMelt-Bet 
	as shown in Figure \ref{Fig_event_prop_control}. This implies that the proposed LES is indeed more effective in identifying most important follower links that influence event propagation.

	\section{Conclusion and Future Work}
	\label{sec_con}
	The contributions of this paper are twofold. First, we have
exploited the user languages on Twitter to  discover the Network of Networks (NoN) structure embedded in real-world event propagation patterns. 
	Second, by minimizing an upper bound on event propagation increment, a left eigenvector score (LES) is proposed to identify the level of importance of a link in event propagation, which is the product of eigenvector centrality of the associated node pair.
Experiments on synthetic datasets demonstrate the advantage of incorporating the NoN structure for identifying influential links. Moreover,
	 experiments on Twitter data show that the proposed method is able to exploit the NoN structure over the different languages used by Twitter users. 
	In particular, we show that the 
	LES successfully identifies most important links influencing event propagation. Potential future directions are the incorporation of multiple NoN models on a single network for identifying influential links and for other network analysis and applications.

	%\section*{Acknowledgment}
	%The authors would like to thank Baichuan Zhang at the Department of Computer and Information Science, Indiana University - Purdue University Indianapolis, for his help in collecting the Leskovec-Ng collaboration network dataset and producing the figure in Fig. \ref{Fig_JureNg}.
	
	\bibliographystyle{IEEEtran}
	\bibliography{IEEEabrv20160824,CPY_ref_20171130}

\begin{thebibliography}{10}
\providecommand{\url}[1]{#1}
\csname url@rmstyle\endcsname
\providecommand{\newblock}{\relax}
\providecommand{\bibinfo}[2]{#2}
\providecommand\BIBentrySTDinterwordspacing{\spaceskip=0pt\relax}
\providecommand\BIBentryALTinterwordstretchfactor{4}
\providecommand\BIBentryALTinterwordspacing{\spaceskip=\fontdimen2\font plus
\BIBentryALTinterwordstretchfactor\fontdimen3\font minus
  \fontdimen4\font\relax}
\providecommand\BIBforeignlanguage[2]{{%
\expandafter\ifx\csname l@#1\endcsname\relax
\typeout{** WARNING: IEEEtran.bst: No hyphenation pattern has been}%
\typeout{** loaded for the language `#1'. Using the pattern for}%
\typeout{** the default language instead.}%
\else
\language=\csname l@#1\endcsname
\fi
#2}}

\bibitem{Pastor-Satorras15epidemic_review}
R.~Pastor-Satorras, C.~Castellano, P.~Van~Mieghem, and A.~Vespignani,
  ``Epidemic processes in complex networks,'' \emph{Rev. Mod. Phys.}, vol.~87,
  pp. 925--979, Aug 2015.

\bibitem{nowzari2016analysis}
C.~Nowzari, V.~M. Preciado, and G.~J. Pappas, ``Analysis and control of
  epidemics: A survey of spreading processes on complex networks,''
  \emph{{IEEE} Control Syst. Mag.}, vol.~36, no.~1, pp. 26--46, 2016.

\bibitem{yang2010predicting}
J.~Yang and S.~Counts, ``Predicting the speed, scale, and range of information
  diffusion in twitter,'' \emph{International Conference on Web and Social
  Media}, vol.~10, pp. 355--358, 2010.

\bibitem{kitsak2010identification}
M.~Kitsak, L.~K. Gallos, S.~Havlin, F.~Liljeros, L.~Muchnik, H.~E. Stanley, and
  H.~A. Makse, ``Identification of influential spreaders in complex networks,''
  \emph{Nature Physics}, vol.~6, no.~11, pp. 888--893, 2010.

\bibitem{myers2012information}
S.~A. Myers, C.~Zhu, and J.~Leskovec, ``Information diffusion and external
  influence in networks,'' in \emph{ACM International Conference on Knowledge
  Discovery and Data Mining (KDD)}.\hskip 1em plus 0.5em minus 0.4em\relax ACM,
  2012, pp. 33--41.

\bibitem{Arruda14}
G.~F. de~Arruda, A.~L. Barbieri, P.~M. Rodr\'{\i}guez, F.~A. Rodrigues,
  Y.~Moreno, and L.~d.~F. Costa, ``Role of centrality for the identification of
  influential spreaders in complex networks,'' \emph{Phys. Rev. E}, vol.~90, p.
  032812, Sep 2014.

\bibitem{kimura2009blocking}
M.~Kimura, K.~Saito, and H.~Motoda, ``Blocking links to minimize contamination
  spread in a social network,'' \emph{ACM Transactions on Knowledge Discovery
  from Data (TKDD)}, vol.~3, no.~2, p.~9, 2009.

\bibitem{del2016spreading}
M.~Del~Vicario, A.~Bessi, F.~Zollo, F.~Petroni, A.~Scala, G.~Caldarelli, H.~E.
  Stanley, and W.~Quattrociocchi, ``The spreading of misinformation online,''
  \emph{Proceedings of the National Academy of Sciences}, vol. 113, no.~3, pp.
  554--559, 2016.

\bibitem{Radicchi16influential}
F.~Radicchi and C.~Castellano, ``Leveraging percolation theory to single out
  influential spreaders in networks,'' \emph{Phys. Rev. E}, vol.~93, p. 062314,
  Jun 2016.

\bibitem{Cohen03immu}
R.~Cohen, S.~Havlin, and D.~ben Avraham, ``Efficient immunization strategies
  for computer networks and populations,'' \emph{Phys. Rev. Lett.}, vol.~91, p.
  247901, Dec 2003.

\bibitem{zou2007modeling}
C.~C. Zou, D.~Towsley, and W.~Gong, ``Modeling and simulation study of the
  propagation and defense of internet e-mail worms,'' \emph{{IEEE} Trans.
  Depend. Sec. Comput.}, vol.~4, no.~2, pp. 105--118, 2007.

\bibitem{Chen08immu}
Y.~Chen, G.~Paul, S.~Havlin, F.~Liljeros, and H.~E. Stanley, ``Finding a better
  immunization strategy,'' \emph{Phys. Rev. Lett.}, vol. 101, p. 058701, Jul
  2008.

\bibitem{gao2013modeling}
C.~Gao and J.~Liu, ``Modeling and restraining mobile virus propagation,''
  \emph{{IEEE} Trans. Mobile Comput.}, vol.~12, no.~3, pp. 529--541, 2013.

\bibitem{de2013anatomy}
M.~De~Domenico, A.~Lima, P.~Mougel, and M.~Musolesi, ``The anatomy of a
  scientific rumor,'' \emph{Scientific reports}, vol.~3, 2013.

\bibitem{Gao11NoN}
J.~Gao, S.~V. Buldyrev, S.~Havlin, and H.~E. Stanley, ``Robustness of a network
  of networks,'' \emph{Phys. Rev. Lett.}, vol. 107, p. 195701, Nov 2011.

\bibitem{Roni10}
R.~Parshani, S.~V. Buldyrev, and S.~Havlin, ``Interdependent networks: Reducing
  the coupling strength leads to a change from a first to second order
  percolation transition,'' \emph{Phys. Rev. Lett.}, vol. 105, p. 048701, Jul
  2010.

\bibitem{buldyrev2010catastrophic}
S.~V. Buldyrev, R.~Parshani, G.~Paul, H.~E. Stanley, and S.~Havlin,
  ``Catastrophic cascade of failures in interdependent networks,''
  \emph{Nature}, vol. 464, no. 7291, pp. 1025--1028, 2010.

\bibitem{Anna12}
A.~Saumell-Mendiola, M.~A. Serrano, and M.~Bogu\~n\'a, ``Epidemic spreading on
  interconnected networks,'' \emph{Phys. Rev. E}, vol.~86, p. 026106, Aug 2012.

\bibitem{Radicchi13}
F.~Radicchi and A.~Arenas, ``Abrupt transition in the structural formation of
  interconnected networks,'' \emph{Nature Physics}, vol.~9, no.~11, pp.
  717--720, Nov. 2013.

\bibitem{ni2014inside}
J.~Ni, H.~Tong, W.~Fan, and X.~Zhang, ``Inside the atoms: ranking on a network
  of networks,'' in \emph{ACM International Conference on Knowledge Discovery
  and Data Mining (KDD)}, 2014, pp. 1356--1365.

\bibitem{Zou05}
C.~Zou, W.~Gong, D.~Towsley, and L.~Gao, ``The monitoring and early detection
  of {Internet} worms,'' \emph{{IEEE/ACM} Trans. Netw.}, vol.~13, no.~5, pp.
  961--974, Oct. 2005.

\bibitem{CPY14control}
P.-Y. Chen, S.-M. Cheng, and K.-C. Chen, ``Optimal control of epidemic
  information dissemination over networks,'' \emph{IEEE Trans. on Cybern.},
  vol.~44, no.~12, pp. 2316--2328, Dec. 2014.

\bibitem{cha2009measurement}
M.~Cha, A.~Mislove, and K.~P. Gummadi, ``A measurement-driven analysis of
  information propagation in the flickr social network,'' in \emph{ACM
  International conference on world wide web (WWW)}, 2009, pp. 721--730.

\bibitem{kempe2003maximizing}
D.~Kempe, J.~Kleinberg, and {\'E}.~Tardos, ``Maximizing the spread of influence
  through a social network,'' in \emph{ACM International Conference on
  Knowledge Discovery and Data Mining (KDD)}.\hskip 1em plus 0.5em minus
  0.4em\relax ACM, 2003, pp. 137--146.

\bibitem{shah2011rumors}
D.~Shah and T.~Zaman, ``Rumors in a network: who's the culprit?'' \emph{{IEEE}
  Trans. Inf. Theory}, vol.~57, no.~8, pp. 5163--5181, 2011.

\bibitem{lopes2008diffusion}
C.~G. Lopes and A.~H. Sayed, ``Diffusion least-mean squares over adaptive
  networks: Formulation and performance analysis,'' \emph{{IEEE} Trans. Signal
  Process.}, vol.~56, no.~7, pp. 3122--3136, 2008.

\bibitem{Gholami16}
M.~R. Gholami, M.~Jansson, E.~G. Ström, and A.~H. Sayed, ``Diffusion
  estimation over cooperative multi-agent networks with missing data,''
  \emph{{IEEE} Trans. Signal Inf. Process. Netw.}, vol.~PP, no.~99, pp. 1--1,
  2016.

\bibitem{benzi2016principal}
K.~Benzi, B.~Ricaud, and P.~Vandergheynst, ``Principal patterns on graphs:
  Discovering coherent structures in datasets,'' \emph{{IEEE} Trans. Signal
  Inf. Process. Netw.}, vol.~2, no.~2, pp. 160--173, 2016.

\bibitem{hamon2016extraction}
R.~Hamon, P.~Borgnat, P.~Flandrin, and C.~Robardet, ``Extraction of temporal
  network structures from graph-based signals,'' \emph{{IEEE} Trans. Signal
  Inf. Process. Netw.}, vol.~2, no.~2, pp. 215--226, 2016.

\bibitem{segarra2015diffusion}
S.~Segarra, W.~Huang, and A.~Ribeiro, ``Diffusion and superposition distances
  for signals supported on networks,'' \emph{{IEEE} Trans. Signal Inf. Process.
  Netw.}, vol.~1, no.~1, pp. 20--32, 2015.

\bibitem{goldenberg2001using}
J.~Goldenberg, B.~Libai, and E.~Muller, ``Using complex systems analysis to
  advance marketing theory development: Modeling heterogeneity effects on new
  product growth through stochastic cellular automata,'' \emph{Academy of
  Marketing Science Review}, vol. 2001, p.~1, 2001.

\bibitem{tong2012gelling}
H.~Tong, B.~A. Prakash, T.~Eliassi-Rad, M.~Faloutsos, and C.~Faloutsos,
  ``Gelling, and melting, large graphs by edge manipulation,'' in \emph{ACM
  International Conference on Information and Knowledge Management (CIKM)},
  2012, pp. 245--254.

\bibitem{silver2016mastering}
D.~Silver, A.~Huang, C.~J. Maddison, A.~Guez, L.~Sifre, G.~Van Den~Driessche,
  J.~Schrittwieser, I.~Antonoglou, V.~Panneershelvam, M.~Lanctot,
  \emph{et~al.}, ``Mastering the game of go with deep neural networks and tree
  search,'' \emph{Nature}, vol. 529, no. 7587, pp. 484--489, 2016.

\bibitem{prakash2012threshold}
B.~A. Prakash, D.~Chakrabarti, N.~C. Valler, M.~Faloutsos, and C.~Faloutsos,
  ``Threshold conditions for arbitrary cascade models on arbitrary networks,''
  \emph{Knowledge and Information Systems}, vol.~33, no.~3, pp. 549--575, 2012.

\bibitem{Deri_spectral17}
J.~A. Deri and J.~M.~F. Moura, ``Spectral projector-based graph fourier
  transforms,'' \emph{{IEEE} J. Sel. Topics Signal Process.}, vol.~11, no.~6,
  pp. 785--795, Sept. 2017.

\bibitem{HornMatrixAnalysis}
R.~A. Horn and C.~R. Johnson, \emph{{Matrix Analysis}}.\hskip 1em plus 0.5em
  minus 0.4em\relax Cambridge University Press, 1990.

\bibitem{Newman10NetworkIntro}
M.~E.~J. Newman, \emph{Networks: An Introduction}.\hskip 1em plus 0.5em minus
  0.4em\relax Oxford University Press, Inc., 2010.

\bibitem{chen2010scalable}
W.~Chen, C.~Wang, and Y.~Wang, ``Scalable influence maximization for prevalent
  viral marketing in large-scale social networks,'' in \emph{ACM International
  Conference on Knowledge Discovery and Data Mining (KDD)}, 2010, pp.
  1029--1038.

\bibitem{Girvan02}
M.~Girvan and M.~E.~J. Newman, ``Community structure in social and biological
  networks,'' \emph{Proc. National Academy of Sciences}, vol.~99, no.~12, pp.
  7821--7826, 2002.

\bibitem{johnson1977efficient}
D.~B. Johnson, ``Efficient algorithms for shortest paths in sparse networks,''
  \emph{Journal of the ACM}, vol.~24, no.~1, pp. 1--13, 1977.

\bibitem{easley2010networks}
D.~Easley and J.~Kleinberg, \emph{Networks, Crowds, and Markets: Reasoning
  About A Highly Connected World}.\hskip 1em plus 0.5em minus 0.4em\relax
  Cambridge University Press, 2010.

\bibitem{tang2012inferring}
J.~Tang, T.~Lou, and J.~Kleinberg, ``Inferring social ties across heterogenous
  networks,'' in \emph{ACM International Conference on Web Search and Data
  Mining}, 2012, pp. 743--752.

\bibitem{meyer2000matrix}
C.~D. Meyer, \emph{Matrix analysis and applied linear algebra}.\hskip 1em plus
  0.5em minus 0.4em\relax Siam, 2000, vol.~2.

\end{thebibliography}
	
\appendix
\subsection{Details of the collected real-world event propagation traces on Twitter}
\label{appen_detail_NON}
To illustrate event propagation, we collected the traces of three recent events on Twitter\footnote{Datasets available at \url{https://sites.google.com/site/pinyuchenpage/datasets}} during a period of two weeks through the Twitter API. These events include URLs and hashtags specified as follows.~\\

\begin{itemize}
	\item \textbf{Obama FB:} we tracked the tweets including the URL ``http://Facebook.com/POTUS'' from November 9th to November 23rd in 2015. The URL links to U.S. President Obama's personal Facebook page, and was firstly being posted by his personal Twitter account on November 9th 2015.
	\item \textbf{Premier 12:} we tracked the tweets including the hashtag ``\#premier12'' from November 19th to December 3rd in 2015. Premier 12 is a flagship international baseball tournament organized by the World Baseball Softball Confederation (WBSC), featuring the twelve best-ranked national baseball teams in the world.
	\item \textbf{AlphaGo:}	we tracked the tweets including the hashtag ``\#AlphaGo'' from January 27th to February 10th in 2016. AlphaGo is a computer program developed by Google DeepMind in London to play the board game Go. On January 27th 2016, the news of AlphaGo defeating a European Go champion was announced along with the algorithm published in Nature \cite{silver2016mastering}.
\end{itemize}

%\subsection{Derivation of the iterative state equation in (\ref{eqn_retweet_propagation_model})}
%\label{appen_iterative_state}
%%First, recall that the definition of the entry-wise threshold function $\TB(\cdot)$ is for any non-negative vector $\bx$, $[\TB(\bx)]_i=1$ if $[\bx]_i>1$ and $[\TB(\bx)]_i=[\bx]_i$ if $0 \leq [\bx]_i \leq 1$. It is easy to show that for any two nonnegative vectors $\bx_1$ and $\bx_2$ of the same dimension, 
%%\begin{align}
%%\label{eqn_binary_threshold_prop_1}
%%\TB(\bx_1+\bx_2)=\TB(\bx_1+\TB(\bx_2)).
%%\end{align}
%
%Since $\bA_t$ accounts for the adjacency matrix of activated follower links for event propagation during the $t$-th time frame, the $i$-th entry of the vector $\bA_{t+1}^T \br_t$ can be expressed as $[\bA_{t+1}^T \br_t]_i=\sum_{j=1}^n [\bA_t]_{ij} [\br_t]_j$, which is the number of tweets regarding the event that user $i$ decides to share on Twitter during the $t+1$-th time frame. Therefore, the entry-wise thresholded binary vector $\TB(\bA_{t+1}^T \br_t)$  indicates the status of new users participating in event propagation during the $t+1$-th time frame. Lastly, since $\TB(\bA_{t+1}^T \br_t)$ represents the vector of event propagation increment, $\br_{t+1}=\TB(\br_t+\TB(\bA_{t+1}^T \br_t))$ accounts for the event propagation status of all users since the beginning to the $t+1$-th time frame.

\subsection{Proof of the upper bound in (\ref{eqn_surrogate_retweet})}
\label{appen_UB}
First, observe from (\ref{eqn_retweet_propagation_model}) that the sparsity level $\| \br_t \|_0$ of $\br_t$ is a non-decreasing function in $t$. Therefore, the condition that $\|\br_F\|_0 \leq s$ implies $\|\br_t\|_0 \leq s$ for all $t \leq F$. Let $\bone_n$ denote the $n$-dimensional column vector of ones. Then the sparsity level $\|\TB(\bA_{t+1} \br_t)\|_0$ of the binary vector $\TB(\bA_{t+1} \br_t)$ can be expressed as 
\begin{align}
\label{eqn_sparsity_level_eqiv}
\|\TB(\bA_{t+1} \br_t)\|_0=\bone_n^T \TB(\bA_{t+1} \br_t),
\end{align}
where $\cdot^T$ denotes the transpose of a matrix (or a vector).

Let $\|\bx\|_2=\lb \sum_{i=1}^n [\bx]_i^2 \rb^{1/2}$ denote the Euclidean norm of a vector $\bx$ and let $\bA_{t+1}=\bV \bSigma \bV^{-1} $ be the eigendecomposition of $\bA_{t+1}$, where $\bV$ is the matrix whose columns are the right eigenvectors of   $\bA_{t+1}$ and  $\bSigma$ is the diagonal matrix whose diagonal elements are the corresponding eigenvalues.
We can derive an upper bound on the term $\bone_n^T \TB(\bA_{t+1} \br_t)$ in (\ref{eqn_sparsity_level_eqiv}), which is
\begin{align}
\label{eqn_NoN_upper_bound_proof_2}
&\bone_n^T \TB(\bA_{t+1} \br_t)  \nonumber \\
&\overset{(a)}{\leq} \bone_n^T \bA_{t+1} \br_t  \nonumber \\
&\overset{(b)}{=} \bone_n^T \bV \bSigma \bV^{-1}  \br_t   \nonumber \\
&\overset{(c)}{=} \|\bone_n^T \bV\|_2 \cdot \|\bV^{-1}  \br_t \|_2 \cdot   \frac{\bone_n^T \bV}{\|\bone_n^T \bV\|_2} \bSigma \frac{\bV^{-1}  \br_t}{\|\bV^{-1}  \br_t \|_2}   \nonumber\\
&\overset{(d)}{\leq}   \|\bone_n^T \bV\|_2 \cdot \|\bV^{-1}  \br_t \|_2 \cdot \max_{\bx,\bz:\|\bx\|_2=1, \|\bz\|_2=1} \bx^T  \bSigma \bz  \nonumber \\
&\overset{(e)}{\leq}   \|\bone_n^T \bV\|_2 \cdot \|\bV^{-1}  \br_t \|_2 \cdot \textnormal{rank}(\bSigma) \cdot \lambda_{\max} (\bA_{t+1})   \nonumber \\
&\overset{(f)}{\leq}  \|\bone_n^T \bV\|_2  \cdot  \sqrt{s} \cdot  \|\bV^{-1}\|_{\textnormal{op}} \cdot \textnormal{rank}(\bSigma)\cdot \lambda_{\max} (\bA_{t+1})   \nonumber \\
&\overset{(g)}{\leq} C_{t+1} \sqrt{s} \cdot \lambda_{\max} (\bA),
\end{align}
where $ C_{t+1}=\|\bone_n^T \bV\|_2 \cdot  \|\bV^{-1}\|_{\textnormal{op}} \cdot \textnormal{rank}(\bSigma)$ and $\|\bM\|_{\textnormal{op}}=\sup_{\bx:\|\bx\|_2=1} \|\bM \bx \|_2$ denotes the operator norm  of a square matrix $\bM$ induced by $\|\cdot\|_2$ (also known as the spectral norm).
The inequality in $(a)$ follows from the facts that $\TB(\cdot)$ is a threshold function and that $\bA_{t+1} \br_t$ is a nonnegative vector. The equality in $(b)$ holds by substituting $\bA_{t+1}$ with its eigendecomposition. The equality in $(c)$ is a simple arithmetic operation. The inequality in $(d)$ holds by taking the maximum value of the term $\bx^T  \bSigma \bz$ with unit-norm constraints on $\bx$ and $\bz$. The inequality in $(e)$ is due to the fact that $\max_{\bx,\bz:\|\bx\|_2=1, \|\bz\|_2=1} \bx^T  \bSigma \bz \leq \textnormal{rank}(\bSigma) \cdot \max_{i \in \{1,\ldots,n\} } |[\bSigma]_{ii}| 
\leq \textnormal{rank}(\bSigma) \cdot \lambda_{\max} (\bA_{t+1})  $ by the Perron-Frobenius theorem \cite{HornMatrixAnalysis}. Note that if $\bA_{t+1}$ is a reducible matrix, one can use its equivalent upper-triangular block form and the fact the Perron root (i.e., $\lambda_{\max} (\bA_{t+1})$ is the maximum of the spectral radius of the irreducible blocks \cite{meyer2000matrix}. 
The inequality in $(f)$ follows by $\|\bV^{-1}  \br_t \|_2 \leq \|\bV^{-1}\|_{\textnormal{op}} \cdot \|\br_t\|_2$ and $\|\br_t\|_2^2=\sum_{i=1}^n [\br_t]_i^2=\sum_{i=1}^n [\br_t]_i=\| \br_t \|_0 \leq s$ since $\br_t$ is a binary vector. The inequality in $(g)$ holds because by definition $\bA-\bA_{t+1}$ is also a nonnegative matrix \cite[Chapter~7 and 8]{meyer2000matrix}.

Lastly, using (\ref{eqn_NoN_upper_bound_proof_2}), take $C=\max_{t \in \{0,\ldots,F-1\}} C_{t+1}$. Then we obtain the upper bound on $\|\TB(\bA_{t+1} \br_t)\|_0$ for all $t \in \{0,\ldots,F-1\}$ as stated in (\ref{eqn_surrogate_retweet}).

\subsection{Proof of the bounds in (\ref{eqn_event_propagation_bound_general}), (\ref{eqn_event_propagation_bound}) and (\ref{eqn_event_propagation_bound_2})}
\label{appen_UB_2}
Given a follower link removal set $\cE_{\cR}$ with cardinality $|\cE_{\cR}|=q \geq 1$, the adjacency matrix $\bAt(\cE_{\cR})$ after removing the follower links in $\cE_{\cR}$ from the original network can be written as a matrix perturbation to the adjacency matrix $\bA$ of the original Twitter follower network, which takes the form
\begin{align}
\label{eqn_dummy_1}
\bAt(\cE_{\cR})=\bA - \sum_{(i,j) \in \cE_{\cR}} \be_i \be_j^T,
\end{align}
where $\be_i$ denotes the $n$-dimensional column vector of zeros except that its $i$-th entry is $1$.

To obtain a lower bound on $ \lambda_{\max} (\bAt(\cE_{\cR}))$ in terms of $\lambda_{\max} (\bA)$ and $\sum_{(i,j) \in \cE_{\cR}} [\by]_i [\by]_j$, we denote the eigendecomposition of $\bAt(\cE_{\cR})$  by $\bAt(\cE_{\cR})=\bV_{\bAt} \bSigma_{\bAt} \bV_{\bAt}^{-1}$.
	 Note that
	for any nonnegative unit-length vector $\bx \in \bbR^{n}$  and any scalar $b$, 
\begin{align}
\label{eqn_dummy_new_0}
&(b \bx + \by)^T \bAt(\cE_{\cR}) (b \bx + \by)  \nonumber \\
&= \|b \bx + \by\|_2^2 \cdot \frac{(b \bx + \by)^T}{\|b \bx + \by\|_2} \bAt(\cE_{\cR}) \frac{(b \bx + \by)}{\|b \bx + \by\|_2} \nonumber \\
&\overset{(i)}{\leq} (1+b)^2 \cdot \frac{(b \bx + \by)^T}{\|b \bx + \by\|_2} \bAt(\cE_{\cR}) \frac{(b \bx + \by)}{\|b \bx + \by\|_2} \nonumber \\
&\overset{(ii)}{\leq} (1+b)^2 \cdot K_1 \cdot \lambda_{\max}(\bAt(\cE_{\cR})),
\end{align}	 
where $(i)$ is due to the fact that $\|b \bx + \by\|_2^2=1+b^2+2b\bx^T \by \leq 1+b^2+2b \|\bx\|_2 \|\byt\|_2=(1+b)^2$ by the Cauchy-Schwartz inequality, and $K_1=\|\bV_{\bAt}\|_{\textnormal{op}}  \cdot  \|\bV_{\bAt}^{-1}\|_{\textnormal{op}} \cdot \textnormal{rank}(\bSigma_{\bAt})$.
 The inequality in $(ii)$ holds because
 \begin{align}
% \label{eqn_}
   &\frac{(b \bx + \by)^T}{\|b \bx + \by\|_2} \bAt(\cE_{\cR}) \frac{(b \bx + \by)}{\|b \bx + \by\|_2} \nonumber \\
   & = \frac{(b \bx + \by)^T}{\|b \bx + \by\|_2} \bV_{\bAt} \bSigma_{\bAt} 
   \bV_{\bAt}^{-1} \frac{(b \bx + \by)}{\|b \bx + \by\|_2} \nonumber \\ 
   & \leq \|\frac{(b \bx + \by)^T}{\|b \bx + \by\|_2} \bV_{\bAt}\|_2 \cdot \|  \bV_{\bAt}^{-1} \frac{(b \bx + \by)}{\|b \bx + \by\|_2}\|_2 \cdot
   \max_{\bz:\|\bz\|_2=1} \bz^T \bSigma_{\bAt} \bz  \nonumber \\
   &  \overset{(iii)}{\leq} \|\bV_{\bAt}\|_{\textnormal{op}} \cdot  \|\bV_{\bAt}^{-1}\|_{\textnormal{op}}  \cdot \max_{\bz:\|\bz\|_2=1} \bz^T \bSigma_{\bAt} \bz  \nonumber \\
   &  \overset{(iv)}{\leq} \|\bV_{\bAt}\|_{\textnormal{op}} \cdot  \|\bV_{\bAt}^{-1}\|_{\textnormal{op}}  \cdot \textnormal{rank}(\bSigma_{\bAt}) \cdot \lambda_{\max}(\bAt(\cE_{\cR})),
 \end{align}
 where the inequality in $(iii)$ follows by $\|\frac{(b \bx + \by)^T}{\|b \bx + \by\|_2} \bV_{\bAt}\|_2 \leq \|\bV_{\bAt}\|_{\textnormal{op}} $ and  $ \|  \bV_{\bAt}^{-1} \frac{(b \bx + \by)}{\|b \bx + \by\|_2}\|_2 \leq \|\bV_{\bAt}^{-1}\|_{\textnormal{op}} $, and the inequality in $(iv)$ is due to the fact that $\max_{\bz: \|\bz\|_2=1} \bz^T \bSigma_{\bAt} \bz \leq \textnormal{rank}(\bSigma_{\bAt}) \cdot \max_{i \in \{1,\ldots,n\} } |[\bSigma_{\bAt}]_{ii}| 
 \leq \textnormal{rank}(\bSigma_{\bAt}) \cdot \lambda_{\max}(\bAt(\cE_{\cR}))  $ by the Perron-Frobenius theorem \cite{HornMatrixAnalysis}. If $\bAt(\cE_{\cR})$ is a reducible matrix, one can use its equivalent upper-triangular block form and the fact the Perron root (i.e., $\lambda_{\max} (\bAt(\cE_{\cR}))$ is the maximum of the spectral radius of the irreducible blocks \cite{meyer2000matrix}.

Expanding  the left hand side (LHS) of (\ref{eqn_dummy_new_0}) and using (\ref{eqn_dummy_1}) to replace $\bAt(\cE_{\cR})$ by $\bA - \sum_{(i,j) \in \cE_{\cR}} \be_i \be_j^T$, we have 
\begin{align}
\label{eqn_dummy_new_01}
&(b \bx + \by)^T \bAt(\cE_{\cR}) (b \bx + \by)   \nonumber \\
&= \lambda_{\max}(\bA) -   \sum_{(i,j) \in \cE_{\cR}} [\by]_i [\by]_j^T + b^2 \bx^T \bAt(\cE_{\cR}) \bx
+ b \bx^T \bAt(\cE_{\cR}) \by \nonumber \\
&~~~ + b \by^T \bAt(\cE_{\cR}) \bx.
\end{align}	 
Recalling that $\bA$ and $\bAt(\cE_{\cR})$ are binary matrices, and
$\bx$ and $\by$ are nonnegative vectors, 	we define
\begin{align}
\label{eqn_dummy_new_8} 
 b^2 \bx^T \bAt(\cE_{\cR}) \bx
 + b \bx^T \bAt(\cE_{\cR}) \by + b \bx^T \bAt(\cE_{\cR}) \by =a  \lambda_{\max}  (\bAt(\cE_{\cR}))
\end{align}
for some $a$. Then we can rewrite (\ref{eqn_dummy_new_0}) as 
\begin{align}
\label{eqn_dummy_new_9}
& \lambda_{\max}(\bA) -   \sum_{(i,j) \in \cE_{\cR}} [\by]_i [\by]_j^T \leq [(1+b)^2  K_1 -a] \cdot \lambda_{\max}(\bAt(\cE_{\cR})).
\end{align}	  
Setting $(1+b)^2  K_1  = a+1$ gives the lower bound on $ \lambda_{\max} (\bAt(\cE_{\cR}))$ as stated in (\ref{eqn_event_propagation_bound}).

Next, we use similar analysis technique to obtain an upper bound on $ \lambda_{\max} (\bAt(\cE_{\cR}))$ in terms of $\lambda_{\max} (\bA)$ and $\sum_{(i,j) \in \cE_{\cR}} [\bx]_i [\bx]_j$. We denote the eigendecomposition of $\bA$  by $\bA=\bV_{\bA} \bSigma_{\bA} \bV_{\bA}^{-1}$, and let $\byt$ denote the leading left eigenvector of $\bAt(\cE_{\cR})$. Note that
for any nonnegative unit-length vector $\bx \in \bbR^{n}$  and any scalar $b$, 
\begin{align}
\label{eqn_dummy_new_1}
(b \bx + \byt)^T \bA (b \bx + \byt) &= \|b \bx + \byt\|_2^2 \cdot \frac{(b \bx + \byt)^T}{\|b \bx + \byt\|_2} \bA \frac{(b \bx + \byt)}{\|b \bx + \byt\|_2} \nonumber \\
&\overset{(v)}{\leq} (1+b)^2 \cdot \frac{(b \bx + \byt)^T}{\|b \bx + \byt\|_2} \bA \frac{(b \bx + \byt)}{\|b \bx + \byt\|_2} \nonumber \\
&\overset{(vi)}{\leq} (1+b)^2 \cdot K_2 \cdot \lambda_{\max}(\bA),
\end{align}	 
where $K_2=\|\bV_{\bA}\|_{\textnormal{op}}  \cdot  \|\bV_{\bA}^{-1}\|_{\textnormal{op}} \cdot \textnormal{rank}(\bSigma_{\bA})$. The inequalities in $(v)$ and $(vi)$ of $(\ref{eqn_dummy_new_1})$ are similar to the inequalities in $(i)$ and $(ii)$ of $(\ref{eqn_dummy_new_0})$.

Expanding the LHS of (\ref{eqn_dummy_new_1}) and Using (\ref{eqn_dummy_1}) to replace $\bA$ by $\bAt(\cE_{\cR})+ \sum_{(i,j) \in \cE_{\cR}} \be_i \be_j^T$, we have 
\begin{align}
\label{eqn_dummy_new_2}
&(b \bx + \byt)^T \bA (b \bx + \byt)   \nonumber \\
&=b^2 \bx^T \bAt(\cE_{\cR}) \bx + b^2  \sum_{(i,j) \in \cE_{\cR}} [\bx]_i [\bx]_j^T
+ b \bx^T \bA \byt+ b \byt^T \bA \bx  \nonumber \\
&~~~ + \byt^T \bAt(\cE_{\cR}) \byt +  \sum_{(i,j) \in \cE_{\cR}} [\byt]_i [\byt]_j^T \nonumber \\
& \overset{(vii)}{\geq} b^2 \bx^T \bAt(\cE_{\cR}) \bx + b^2 \sum_{(i,j) \in \cE_{\cR}} [\bx]_i [\bx]_j^T + b \bx^T \bA \byt+ b \byt^T \bA \bx   \nonumber \\
&~~~ + \lambda_{\max} (\bAt(\cE_{\cR})) ,
\end{align}	 		
where $(vii)$ follows from the facts that $\byt^T \bAt(\cE_{\cR}) \byt=\lambda_{\max} (\bAt(\cE_{\cR}))$ and  $\sum_{(i,j) \in \cE_{\cR}} [\byt]_i [\byt]_j^T \geq 0$ by the Perron-Frobenius
 theorem \cite{HornMatrixAnalysis}. By defining
 \begin{align}
 \label{eqn_dummy_new_4} 
    b^2 \bx^T \bAt(\cE_{\cR}) \bx  + b \bx^T \bA \byt+ b \byt^T \bA \bx=a  \lambda_{\max}  (\bAt(\cE_{\cR}))
 \end{align}
 for some $a$, the right hand side (RHS) of (\ref{eqn_dummy_new_2}) becomes 
  $(a+1) \lambda_{\max}  (\bAt(\cE_{\cR})) +  b^2 \sum_{(i,j) \in \cE_{\cR}} [\bx]_i [\bx]_j^T$. Comparing to the RHS of (\ref{eqn_dummy_new_1}), we obtain
\begin{align}
\label{eqn_dummy_new_3}
&(a+1)\lambda_{\max} (\bAt(\cE_{\cR})) \leq (1+b)^2 K_2 \lambda_{\max}(\bA) - b^2 \sum_{(i,j) \in \cE_{\cR}} [\bx]_i [\bx]_j^T.
\end{align}	  
 Setting   $a+1=(1+b)^2 K_2$ and dividing both sides by $(1+b)^2 K_2$ in (\ref{eqn_dummy_new_3}) gives 
\begin{align}
\label{eqn_dummy_new_5}
&\lambda_{\max} (\bAt(\cE_{\cR})) \leq \lambda_{\max}(\bA) - c \sum_{(i,j) \in \cE_{\cR}} [\bx]_i [\bx]_j^T, 
\end{align}	  
where $c=\frac{b^2}{(1+b)^2 K_2}$ and  $a+1=(1+b)^2 K_2$, and the result leads to (\ref{eqn_event_propagation_bound_general}).
Finally, setting $\bx=\by$ in (\ref{eqn_dummy_new_5}) gives the upper bound on $ \lambda_{\max} (\bAt(\cE_{\cR}))$ as stated in (\ref{eqn_event_propagation_bound_2}).

\subsection{Implementation of follower link score functions and computational complexity analysis}
\label{appen_implementation}
We consider the score function of a follower link ($i,j$) that takes the form 
\begin{align}
%\label{eqn_score_exp}
\textnormal{score}(i,j)=[\bx]_i \cdot [\bxt]_j, \nonumber
\end{align}
where $\bx$ and $\bxt$ are nonnegative $n$-dimensional vectors. 

The following reports on the implementation and computational complexity of returning $q$ follower links of the highest score for different follower link score functions.

\begin{itemize}
	\item \textbf{LES:} $\bx=\bxt=\by$, where $\by$ is the leading left eigenvector of the adjacency matrix $\bA$. The computational complexity is $O(mq)$, which is analyzed in Sec. \ref{subsec_LES}.
	
	\item \textbf{InDeg:} $\bx=\bxt=\bdin$, where $\bdin$ is the vector of in-degree of each user, and its $j$-th element $[\bdin]_j=\sum_{i=1}^n [\bA]_{ij}$ is the number of followers of user $j$. The computational complexity is $O(mq)$.
	
	\item \textbf{NetMelt:} NetMelt \cite{tong2012gelling} is an edge removal algorithm proposed to decrease the largest eigenvalue $\lambda_{\max}(\bA)$ of the adjacency matrix $\bA$, where $\bx=\by$ and $\bxt=\bz$, and $\bz$ denotes the leading right eigenvector of $\bA$. The computational complexity is $O(mq+n)$.		
			\item \textbf{Edge betweenness:} edge betweenness \cite{Girvan02} requires the information of shortest paths among all node pairs in a network. The importance of an edge is evaluated by	the number of shortest paths that pass through it. Here we use the directed network of the ``followed-by'' information (i.e., the matrix $\bA^T$)  for computing edge betweenness, and set  score$(i,j)$ to be the edge betweenness of the link $(i,j)$.
			The computational complexity of obtaining the shortest paths among all node pairs is $O(nm+n^2 \log n)$ by the Johnson's algorithm \cite{johnson1977efficient}. The overall computation complexity is $O((n+q)m+n^2 \log n)$ for searching the top $q$ links.
		
	\item \textbf{NoN-LES-Bet (NoN-LES-Wit):} NoN-LES-Bet  (NoN-LES-Wit) exploits the NoN structure and evaluates the score function using $\bx=\bxt=\by^{\textnormal{bet}}$ ($\bx=\bxt=\by^{\textnormal{wit}}$), where $\by^{\textnormal{bet}}$ ($\by^{\textnormal{bit}}$) denotes the leading left eigenvector of the between-network (within-network) adjacency matrix $\bAbet$ ($\bAwit$). The computational complexity is $O(mq)$.
	
	\item \textbf{NoN-InDeg-Bet (NoN-InDeg-Wit):} NoN-InDeg-Bet and NoN-InDeg-Wit are extensions of the InDeg score tailored to the NoN structure. Specifically, for NoN-InDeg-Bet (NoN-InDeg-Wit) we set $\bx=\bxt=\bdin^{\textnormal{bet}}$ ( $\bx=\bxt=\bdin^{\textnormal{wit}}$), where $\bdin^\textnormal{bet}$ ($\bdin^\textnormal{wit}$) is the in-degree vector that only accounts for the between-network (within-network) follower links in the Twitter follower network.
	The computational complexity is $O(mq)$.

	\item \textbf{NoN-NetMelt-Bet (NoN-NetMelt-Wit):} Non-NetMelt-Bet and NoN-NetMelt-Wit are NetMelt algorithms that incorporate  the NoN structure. For NoN-Melt-Bet (NoN-NetMelt-Wit), we set $\bx=\by^{\textnormal{bet}}$ and $\bxt=\bz^{\textnormal{bet}}$ ($\bx=\by^{\textnormal{wit}}$ and $\bxt=\bz^{\textnormal{wit}}$), where $\by^{\textnormal{bet}}$ and $\bz^{\textnormal{bet}}$ ($\by^{\textnormal{wit}}$ and $\bz^{\textnormal{wit}}$) denote the left and leading right eigenvectors of $\bAbet$ ($\bAwit$). 
	The computational complexity is $O(mq+n)$.	
\end{itemize}

\end{document}